\documentclass[12pt,a4paper]{article}

\usepackage{amsmath,amssymb}
\usepackage{graphicx}
\usepackage[nosort]{cite}

\usepackage{tikz}
\usetikzlibrary{decorations.markings}
\usepackage{epsfig}
\usepackage{enumerate}
\newcommand{\mathsym}[1]{{}}

\usepackage{amsthm,amsbsy,color,multicol}
\usepackage{array,calc}
\usepackage{rotating}
\usepackage{a4wide,bm}
\usepackage[a4paper,text={450pt,650pt},centering]{geometry}

\let\oldbfseries=\bfseries
\let\oldmdseries=\mdseries
\let\oldnormalfont=\normalfont
\renewcommand{\bfseries}{\oldbfseries\boldmath}
\renewcommand{\mdseries}{\oldmdseries\unboldmath}
\renewcommand{\normalfont}{\oldnormalfont\unboldmath}

\allowdisplaybreaks[3]

\numberwithin{equation}{section}

\usepackage[font=small,labelfont=bf,width=0.85\textwidth]{caption}

\ifx\hypersetup\sadfkjashdfkxja\newcommand\hypersetup[1]{}\fi

\hypersetup{plainpages=false}
\hypersetup{pdfpagemode=UseNone}
\hypersetup{bookmarksnumbered=true}
\hypersetup{pdfstartview=FitH}
\hypersetup{colorlinks=false}
\hypersetup{citebordercolor={.5 1 .5}}
\hypersetup{urlbordercolor={.5 1 1}}
\hypersetup{linkbordercolor={1 .7 .7}}

\DeclareMathSymbol{\Gamma}{\mathalpha}{letters}{"00}
\DeclareMathSymbol{\Delta}{\mathalpha}{letters}{"01}
\DeclareMathSymbol{\Theta}{\mathalpha}{letters}{"02}
\DeclareMathSymbol{\Lambda}{\mathalpha}{letters}{"03}
\DeclareMathSymbol{\Xi}{\mathalpha}{letters}{"04}
\DeclareMathSymbol{\Pi}{\mathalpha}{letters}{"05}
\DeclareMathSymbol{\Sigma}{\mathalpha}{letters}{"06}
\DeclareMathSymbol{\Upsilon}{\mathalpha}{letters}{"07}
\DeclareMathSymbol{\Phi}{\mathalpha}{letters}{"08}
\DeclareMathSymbol{\Psi}{\mathalpha}{letters}{"09}
\DeclareMathSymbol{\Omega}{\mathalpha}{letters}{"0A}

\newcommand{\gen}[1]{\mathrm{#1}}

\newcommand{\dd}{\mathrm{d}}
\newcommand{\ii}{\mathrm{i}}

\ifx\genfrac\sdflkaj\else\fi

\newcommand{\ket}[1]{\left|#1\right\rangle}      
\newcommand{\bra}[1]{\left\langle #1\right|}     

\newcommand{\alg}[1]{\mathfrak{#1}}

\newcommand{\beq}{\begin{equation}}
\newcommand{\eeq}{\end{equation}}

\def\[{\begin{equation}}
\def\]{\end{equation}}
\def\<{\begin{eqnarray}}
\def\>{\end{eqnarray}}

\makeatletter
\def\mr@ignsp#1 {\ifx\:#1\@empty\else #1\expandafter\mr@ignsp\fi}%
\newcommand{\multiref}[1]{\begingroup
\xdef\mr@no@sparg{\expandafter\mr@ignsp#1 \: }%
\def\mr@comma{}%
\@for\mr@refs:=\mr@no@sparg\do{\mr@comma\def\mr@comma{,}\ref{\mr@refs}}%
\endgroup}
\makeatother

\newcommand{\hypref}[2]{\ifx\href\asklfhas #2\else\href{#1}{#2}\fi}

\newcommand{\secref}[1]{Sec.~\multiref{#1}}

\newcommand{\appref}[1]{App.~\multiref{#1}}

\newcommand{\figref}[1]{Fig.~\multiref{#1}}
\renewcommand{\eqref}[1]{(\multiref{#1})}

\makeatletter
\newlength{\apb@width}
\newcommand{\autoparbox}[2][c]{\settowidth{\apb@width}{#2}\parbox[#1]{\apb@width}{#2}}

\makeatother

\ifx\href\asklfhas\newcommand{\href}[2]{#2}\fi

\begin{document}

\renewcommand{\thefootnote}{\fnsymbol{footnote}}
\thispagestyle{empty}
\begin{flushright}\footnotesize
\end{flushright}
\vspace{1cm}

\begin{center}%
{\Large\bfseries%
\hypersetup{pdftitle={Multiple integral representation for the trigonometric SOS model with domain wall boundaries}}%
Multiple integral representation for the \\ trigonometric SOS model with \\ domain wall boundaries%
\par} \vspace{2cm}%

\textsc{W. Galleas}\vspace{5mm}%
\hypersetup{pdfauthor={Wellington Galleas}}%

\textit{ARC Centre of Excellence for the Mathematics  \\ and Statistics of Complex Systems, \\%
The University of Melbourne\\%
VIC 3010, Australia}\vspace{3mm}%

\verb+wgalleas@unimelb.edu.au+ %

\par\vspace{3cm}

\textbf{Abstract}\vspace{7mm}

\begin{minipage}{12.7cm}
Using the dynamical Yang-Baxter algebra we derive a functional equation for the partition
function of the trigonometric SOS model with domain wall boundary conditions. The solution
of the equation is given in terms of a multiple contour integral.

\hypersetup{pdfkeywords={dynamical Yang-Baxter equation, domain wall boundaries, functional equations}}%
\hypersetup{pdfsubject={}}%
\end{minipage}
\vskip 0.1cm
{\small Keywords: Dynamical Yang-Baxter Equation, Functional equations, \\ Domain wall boundaries}
\vskip 2cm
{\small November 2011}

\end{center}

\newpage
\renewcommand{\thefootnote}{\arabic{footnote}}
\setcounter{footnote}{0}

\tableofcontents

\section{Introduction}
\label{sec:intro}

The study of two-dimensional lattice models in Statistical Mechanics advanced dramatically with
the advent of Baxter's concept of commuting transfer matrices \cite{Baxter_1972}. This method introduced the
concept of integrability in Statistical Mechanics and paved the way for the development of a variety 
of exact methods exploring the aforementioned commutativity. As examples of those methods we have
Baxter's $T-Q$ relation \cite{Baxter_1972}, the algebraic Bethe ansatz \cite{Fad_1979}, the inversion trick \cite{Stroganov_1979}, etc. 
Also a variety of models can be tackled by those same techniques such as vertex models, solid-on-solid models
and hard square models. 

Nevertheless, the implementation of those methods depends drastically on the boundary conditions
chosen and the case of domain wall boundaries deserves special attention. This kind of boundary condition for
the six vertex model was introduced by Korepin in \cite{Korepin82} who also obtained a recurrence relation determining the
model partition function. This recurrence relation was later on solved by Izergin in terms of a determinant \cite{Izergin87}.
Moreover, the exact solution of this model raised the issue of the sensitivity of the six vertex model
bulk properties with respect to the
boundary conditions in the thermodynamical limit \cite{Justin_2000}. 

A natural question that emerges in this scenario is how this sensitivity with respect to boundary conditions extends to 
the eight vertex model. In that case, although the partition function has been evaluated in 
\cite{Pakuliak_2008, Rosengren_2008, WengLi_2009}, the lack
of manageable expressions have eluded the analysis of the thermodynamical limit except for 
a particular value of the anisotropy parameter \cite{Rosengren_2011}.

Motivated by this scenario and keeping in mind the relation between Baxter's eight vertex model and solid-on-solid
models, also refereed to as SOS models, here we demonstrate that the algebraic-functional method introduced in
\cite{Galleas10, Galleas11} can also be used in that case. Using that method we derive a  multiple integral formula for
the partition function of the trigonometric SOS model with domain wall boundaries.

This paper is organised as follows. In the \secref{sec:face} we briefly describe 
SOS models in Statistical Mechanics with emphasis on the case of 
domain wall boundary conditions. In the \secref{sec:dyba} we present the
dynamical Yang-Baxter algebra and demonstrate how it can be explored in order to obtain a functional
equation for the partition function of the SOS model with domain wall boundaries. This functional equation is 
analised in \secref{sec:partfun} and the solution of the equation is given in \secref{sec:mint} as a
multiple contour integral. Technical details are presented in \appref{sec:high} through \appref{sec:L1}.

\section{Solid-on-solid (SOS) models}
\label{sec:face}

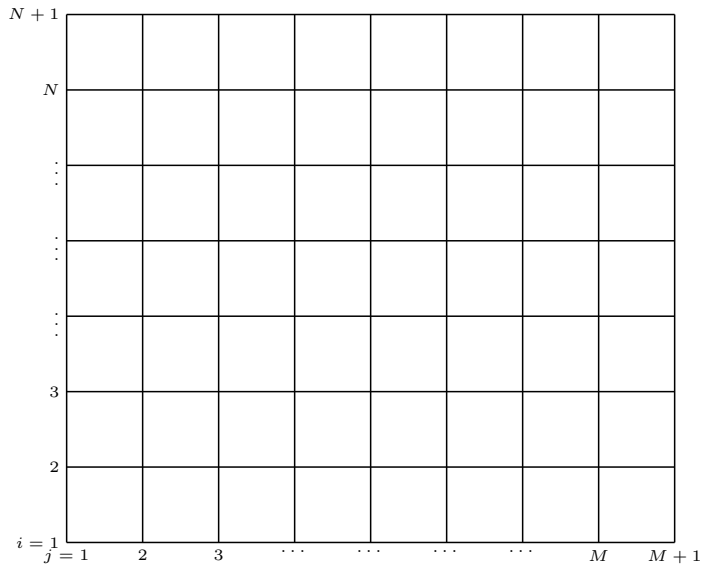
\begin{figure}\centering
\begin{tikzpicture}[scale=0.5,line width=0.2mm]
{\tiny
\draw (0,0) node[left]{$i=1$} -- (16,0) ;
\draw (0,2) node[left]{$2$} -- (16,2) ;
\draw (0,4) node[left]{$3$} -- (16,4) ;
\draw (0,6) node[left]{$\vdots$} -- (16,6) ;
\draw (0,8) node[left]{$\vdots$} -- (16,8) ;
\draw (0,10) node[left]{$\vdots$} -- (16,10) ;
\draw (0,12) node[left]{$N$} -- (16,12) ;
\draw (0,14) node[left]{$N+1$} -- (16,14) ;
\draw (0,0) node[below]{$j=1$} -- (0,14) ;
\draw (2,0) node[below]{$2$} -- (2,14) ;
\draw (4,0) node[below]{$3$} -- (4,14) ;
\draw (6,0) node[below]{$\dots$} -- (6,14) ;
\draw (8,0) node[below]{$\dots$} -- (8,14) ;
\draw (10,0) node[below]{$\dots$} -- (10,14) ;
\draw (12,0) node[below]{$\dots$} -- (12,14) ;
\draw (14,0) node[below]{$M$} -- (14,14) ;
\draw (16,0) node[below]{$M+1$} -- (16,14) ;
}
\end{tikzpicture}
\caption{Two-dimensional lattice with $M \times N$ retangular cells.}
\label{fig:lattice}
\end{figure}

We consider a two-dimesional lattice formed by retangular cells juxtaposed as in \figref{fig:lattice}.
For a lattice with $N+1$ rows and $M+1$ columns we  have $M \times N$ retangular cells and we associate the
Boltzmann weight $w_{ij} \left( \begin{matrix} l_{i+1,j} & l_{i+1,j+1} \cr l_{i,j} & l_{i,j+1} \end{matrix} \right)$
to the cell enclosed by the cartesian coordinates $(i,j)$, $(i,j+1)$, $(i+1,j)$ and $(i+1,j+1)$. 
Each retangular cell is simply referred as {\it face} and its configuration is characterised by the set of variables
$\{ l_{i,j}, l_{i,j+1} , l_{i+1,j} , l_{i+1,j+1} \}$. Diagrammatically this association is depicted in \figref{fig:bw} and
the partition function of the system is then given by
\[
\label{pf}
Z =  \sum_{\{ l_{i,j} \}}  \prod_{i=1}^{N} \prod_{j=1}^{M} w_{ij} \left( \begin{matrix} l_{i+1,j} & l_{i+1,j+1} \cr l_{i,j} & l_{i,j+1} \end{matrix} \right) \; .
\]
\begin{figure} \centering
\begin{tikzpicture}[inner sep=0pt,thick, dot/.style={fill=black,circle,minimum size=5pt}]
\draw (-0.5,0) node[left]{{\large $w_{ij} \left( \begin{matrix} l_{i+1,j} & l_{i+1,j+1} \cr l_{i,j} & l_{i,j+1} \end{matrix} \right) \;\;\; \simeq \;\;\;$}};
    \node[dot] (a) at (0,-1) {};
    \node[dot] (b) at (2,-1) {};
    \node[dot] (c) at (0,1) {};
    \node[dot] (d) at (2,1) {};
\draw (0,-1) -- (2,-1);
\draw (2,-1) -- (2,1);
\draw (2,1) -- (0,1);
\draw (0,1) -- (0,-1);
\draw (0,-1.2) node[below]{{\scriptsize $l_{i,j}$}};
\draw (2,-1.2) node[below]{{\scriptsize  $l_{i,j+1}$}};
\draw (2,1.2) node[above]{{\scriptsize  $l_{i+1,j+1}$}};
\draw (0,1.2) node[above]{{\scriptsize  $l_{i+1,j}$}};
\end{tikzpicture}
\caption{Face at coordinate $(i,j)$ and its Boltzmann weight.}
\label{fig:bw}
\end{figure}

\paragraph{Face monodromy matrix.} Let us define a matrix $\mathcal{T}$ with components
\[
\label{mono}
\mathcal{T}_{l_{i,1}, \dots , l_{i,M+1}}^{l_{i+1,1}, \dots , l_{i+1,M+1}} = \prod_{j=1}^{M} w_{ij} \left( \begin{matrix} l_{i+1,j} & l_{i+1,j+1} \cr l_{i,j} & l_{i,j+1} \end{matrix} \right) \; .
\]   
We shall refer to $\mathcal{T}$ as face monodromy matrix and it is diagrammatically represented in the \figref{fig:mono}.
\begin{figure} \centering
\begin{tikzpicture}[inner sep=0pt,thick, dot/.style={fill=black,circle,minimum size=3pt}]
\draw (-0.5,0.5) node[left]{{\large $\mathcal{T}_{l_{i,1}, \dots , l_{i,M+1}}^{l_{i+1,1}, \dots , l_{i+1,M+1}} \;\;\; \simeq \;\;\;$}};
    \draw[help lines] (0,0) grid (8,1);
\draw (0,-0.2) node[below]{{\scriptsize $l_{i,1}$}};
\draw (1,-0.2) node[below]{{\scriptsize $l_{i,2}$}};
\draw (2,-0.2) node[below]{{\scriptsize $\dots$}};
\draw (6,-0.2) node[below]{{\scriptsize $\dots$}};
\draw (7,-0.2) node[below]{{\scriptsize $\dots$}};
\draw (8,-0.2) node[below]{{\scriptsize $l_{i,M+1}$}};
\draw (0,1.2) node[above]{{\scriptsize $l_{i+1,1}$}};
\draw (1,1.2) node[above]{{\scriptsize $l_{i+1,2}$}};
\draw (2,1.2) node[above]{{\scriptsize $\dots$}};
\draw (6,1.2) node[above]{{\scriptsize $\dots$}};
\draw (7,1.2) node[above]{{\scriptsize $\dots$}};
\draw (8,1.2) node[above]{{\scriptsize $l_{i+1,M+1}$}};
    \node[dot] (a) at (0,0) {};
    \node[dot] (b) at (1,0) {};
    \node[dot] (c) at (2,0) {};
    \node[dot] (d) at (3,0) {};
    \node[dot] (e) at (4,0) {};
    \node[dot] (f) at (5,0) {};
    \node[dot] (g) at (6,0) {};
    \node[dot] (h) at (7,0) {};
    \node[dot] (i) at (8,0) {};
    \node[dot] (j) at (0,1) {};
    \node[dot] (k) at (1,1) {};
    \node[dot] (l) at (2,1) {};
    \node[dot] (m) at (3,1) {};
    \node[dot] (n) at (4,1) {};
    \node[dot] (o) at (5,1) {};
    \node[dot] (p) at (6,1) {};
    \node[dot] (q) at (7,1) {};
    \node[dot] (r) at (8,1) {};
\end{tikzpicture}
\caption{Face monodromy matrix.}
\label{fig:mono}
\end{figure}
For simplicity of notation we also refer to the components of the face monodromy matrix defined by (\ref{mono}) as
$\mathcal{T}_{i}^{i+1}$. Employing this notation the partition function $Z$ reads
\[
\label{pf_mono}
Z = \sum_{\{ l_{i,j} \}} \mathcal{T}_{1}^{2} \mathcal{T}_{2}^{3} \dots \mathcal{T}_{N}^{N+1}  \; ,
\]
and the lattice in \figref{fig:lattice} can be entirely built as the product of elements $\mathcal{T}_{i}^{i+1}$
representing each horizontal layer.

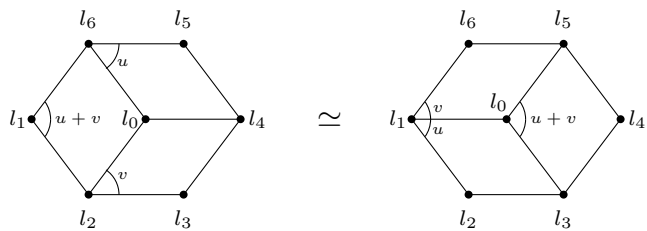
\begin{figure} \centering
\begin{tikzpicture}[inner sep=0pt,thin, dot/.style={fill=black,circle,minimum size=3pt}]
\draw (0,0) -- (-0.75,1) ;
\draw (0,0) -- (0.75,1) ;
\draw (-0.75,1) -- (0,2) ;
\draw (0.75,1) -- (0,2) ;
\draw (0,0) -- (1.25,0) ;
\draw (1.25,0) -- (2,1) ;
\draw (0.75,1) -- (2,1) ;
\draw (0,2) -- (1.25,2) ;
\draw (1.25,2) -- (2,1) ;
    \node[dot] at (0,0) {};
    \node[dot] at (1.25,0) {};
    \node[dot] at (2,1) {};
    \node[dot] at (1.25,2) {};
    \node[dot] at (0,2) {};
    \node[dot] at (-0.75,1) {};
    \node[dot] at (0.75,1) {};
\draw (0.4,2) arc (0:-55:0.4cm);
\draw (0.4,0) arc (0:55:0.4cm);
\draw (-0.5,1) arc (0:-35:0.4cm);
\draw (-0.5,1) arc (0:35:0.4cm);

\draw (-0.45,1) node[right]{{\tiny $u+v$}};
\draw (0.45,1.8) node[below]{{\tiny $u$}};
\draw (0.45,0.2) node[above]{{\tiny $v$}};

\draw (0.68,1) node[left]{{\scriptsize $l_{0}$}};
\draw (-0.8,1) node[left]{{\scriptsize $l_{1}$}};
\draw (2.1,1) node[right]{{\scriptsize $l_{4}$}};
\draw (0,-0.2) node[below]{{\scriptsize $l_{2}$}};
\draw (1.25,-0.2) node[below]{{\scriptsize $l_{3}$}};
\draw (0,2.2) node[above]{{\scriptsize $l_{6}$}};
\draw (1.25,2.2) node[above]{{\scriptsize $l_{5}$}};
\draw (3,1) node[right]{$\simeq$};
\begin{scope}[xshift=5cm]
\draw (0,0) -- (1.25,0) ;
\draw (1.25,0) -- (2,1) ;
\draw (2,1) -- (1.25,2) ;
\draw (1.25,2) -- (0,2) ;
\draw (0,2) -- (-0.75,1) ;
\draw (-0.75,1) -- (0,0) ;
\draw (0.5,1) -- (-0.75,1) ;
\draw (0.5,1) -- (1.25,0) ;
\draw (0.5,1) -- (1.25,2) ;
    \node[dot] at (0,0) {};
    \node[dot] at (1.25,0) {};
    \node[dot] at (2,1) {};
    \node[dot] at (1.25,2) {};
    \node[dot] at (0,2) {};
    \node[dot] at (-0.75,1) {};
    \node[dot] at (0.5,1) {};

\draw (-0.5,1) arc (0:-35:0.4cm);
\draw (-0.5,1) arc (0:35:0.4cm);

\draw (0.75,1) arc (0:-35:0.4cm);
\draw (0.75,1) arc (0:35:0.4cm);

\draw (-0.4,1.2) node[below]{{\tiny $v$}};
\draw (-0.4,0.8) node[above]{{\tiny $u$}};
\draw (1.43,1) node[left]{{\tiny $u+v$}};

\draw (0.4,1.1) node[above]{{\scriptsize $l_{0}$}};
\draw (-0.8,1) node[left]{{\scriptsize $l_{1}$}};
\draw (2.1,1) node[right]{{\scriptsize $l_{4}$}};
\draw (0,-0.2) node[below]{{\scriptsize $l_{2}$}};
\draw (1.25,-0.2) node[below]{{\scriptsize $l_{3}$}};
\draw (0,2.2) node[above]{{\scriptsize $l_{6}$}};
\draw (1.25,2.2) node[above]{{\scriptsize $l_{5}$}};
\end{scope}
\end{tikzpicture}
\caption{Graphical representation of the Yang-Baxter relation for SOS models.}
\label{fig:stface}
\end{figure}

\paragraph{Boundary conditions and integrability.} Following Baxter \cite{Baxter_book} we  consider 
the face monodromy matrix (\ref{mono}) with periodic boundary conditions in the horizontal direction.
This corresponds to setting $l_{i, M+1} = l_{i, 1}$, and under these conditions it is convenient to define
an operator $T$ with components
\[
\label{transfer}
T_{l_{i,1}, \dots , l_{i,M}}^{l_{i+1,1}, \dots , l_{i+1,M}} = w_{i,1} \left( \begin{matrix} l_{i+1,1} & l_{i+1,2} \cr l_{i,1} & l_{i,2} \end{matrix} \right) 
w_{i,2} \left( \begin{matrix} l_{i+1,2} & l_{i+1,3} \cr l_{i,2} & l_{i,3} \end{matrix} \right) \dots w_{i,M} \left( \begin{matrix} l_{i+1,M} & l_{i+1,1} \cr l_{i,M} & l_{i,1} \end{matrix} \right) \; .
\]
The operator $T$ is commonly denominated (face) transfer matrix and it plays an important role in estabilishing the integrability of
two-dimensional lattice models \cite{Baxter_book}. We proceed by defining a second transfer matrix $T'$ similarly to
(\ref{transfer}) but with Boltzmann weights $w'$ instead of $w$. Next we look for conditions on $w$ and $w'$ such that the transfer matrices $T$
and $T'$ form a commutative family, i.e. $\left[ T , T' \right]=0$. This requirement leads to the following relation
\begin{align}
\label{star_triang}
\sum_{l_0} w_{v} \left( \begin{matrix}  l_{2} & l_{0} \cr l_{3} & l_{4} \end{matrix} \right)
w_{u+v} \left( \begin{matrix}  l_{1} & l_{6} \cr l_{2} & l_{0} \end{matrix} \right) &
w_{u} \left( \begin{matrix}  l_{6} & l_{5} \cr l_{0} & l_{4} \end{matrix} \right) \cr 
& = \sum_{l_0} w_{u} \left( \begin{matrix}  l_{1} & l_{0} \cr l_{2} & l_{3} \end{matrix} \right)
w_{u+v} \left( \begin{matrix}  l_{0} & l_{5} \cr l_{3} & l_{4} \end{matrix} \right)
w_{v} \left( \begin{matrix}  l_{1} & l_{6} \cr l_{0} & l_{5} \end{matrix} \right) \; ,  \nonumber \\
\end{align}
where $u$ and $v$ are complex variables parameterising the manifold where the transfer matrices form a commutative family. 
For a detailed derivation of (\ref{star_triang}) we refer to \cite{integrable_book}. The Eq. 
(\ref{star_triang}) is usually referred to as Yang-Baxter relation \cite{Baxter_2010}, or simply 
{\it Hexagon identity}, and the local equivalence transformation described by (\ref{star_triang}) is depicted in \figref{fig:stface}.
Moreover, if we also consider periodic boundary conditions in the vertical direction, i.e. $l_{N+1, j} = l_{1, j}$,
the partition function (\ref{pf}) becomes simply
\[ 
\label{pf_trace}
Z = \mbox{Tr} \left( T^N \right) \; ,
\]
and its evaluation is translated into an eigenvalue problem \cite{Kramers_1941a,Kramers_1941b}. 
Throughout this paper we shall consider a different class of boundary 
conditions where (\ref{pf_trace}) does not apply, although the bulk model is still governed by statistical weights satisfying
(\ref{star_triang}).

\paragraph{The trigonometric SOS model.} The variables $\{ l_{i,j}, l_{i,j+1}, l_{i+1,j} , l_{i+1,j+1} \}$
depicted in \figref{fig:bw} are also called height functions and they characterise the configuration of the   
associated face. The degree of freedom $l_{i,j}$ is also refereed to as the {\it colour} of the face at the 
position $(i,j)$. Furthermore, we can also impose restrictions on $l_{i,j}$ such that only certain configurations
of colours are allowed in the statistical sum (\ref{pf}). 

\begin{figure} \centering
\begin{tikzpicture}[inner sep=0pt,thick, dot/.style={fill=black,circle,minimum size=3pt}, decoration={%
   markings,%
   mark=at position 0.5cm with {\arrow[black]{stealth};}}]

 \draw[help lines] (0,0) grid (2,2);
     \node[dot] (a) at (0,0) {};
     \node[dot] (b) at (1,0) {};
     \node[dot] (c) at (2,0) {};
     \node[dot] (a) at (0,1) {};
     \node[dot] (b) at (1,1) {};
     \node[dot] (c) at (2,1) {};
     \node[dot] (a) at (0,2) {};
     \node[dot] (b) at (1,2) {};
     \node[dot] (c) at (2,2) {};
 \draw (0.4,0.5) node[right]{{\tiny $\theta$}};
 \draw (1.2,0.5) node[right]{{\tiny $\theta+\gamma$}};
 \draw (0.2,1.5) node[right]{{\tiny $\theta-\gamma$}};
 \draw (1.4,1.5) node[right]{{\tiny $\theta$}};
\draw (1.1,-0.5) node[below]{$a_{+}$};
 \begin{scope}[xshift=2.5cm]
 \draw[help lines] (0,0) grid (2,2);
     \node[dot] (a) at (0,0) {};
     \node[dot] (b) at (1,0) {};
     \node[dot] (c) at (2,0) {};
     \node[dot] (a) at (0,1) {};
     \node[dot] (b) at (1,1) {};
     \node[dot] (c) at (2,1) {};
     \node[dot] (a) at (0,2) {};
     \node[dot] (b) at (1,2) {};
     \node[dot] (c) at (2,2) {};
 \draw (0.4,0.5) node[right]{{\tiny $\theta$}};
 \draw (1.2,0.5) node[right]{{\tiny $\theta-\gamma$}};
 \draw (0.2,1.5) node[right]{{\tiny $\theta+\gamma$}};
 \draw (1.4,1.5) node[right]{{\tiny $\theta$}};
\draw (1.1,-0.5) node[below]{$a_{-}$};
 \end{scope}
 \begin{scope}[xshift=5cm]
 \draw[help lines] (0,0) grid (2,2);
     \node[dot] (a) at (0,0) {};
     \node[dot] (b) at (1,0) {};
     \node[dot] (c) at (2,0) {};
     \node[dot] (a) at (0,1) {};
     \node[dot] (b) at (1,1) {};
     \node[dot] (c) at (2,1) {};
     \node[dot] (a) at (0,2) {};
     \node[dot] (b) at (1,2) {};
     \node[dot] (c) at (2,2) {};
 \draw (0.2,0.5) node[right]{{\tiny $\theta+\gamma$}};
 \draw (1.4,0.5) node[right]{{\tiny $\theta$}};
 \draw (0.4,1.5) node[right]{{\tiny $\theta$}};
 \draw (1.2,1.5) node[right]{{\tiny $\theta-\gamma$}};
\draw (1.1,-0.5) node[below]{$b_{+}$};
 \end{scope}
 
 \begin{scope}[xshift=7.5cm]
 \draw[help lines] (0,0) grid (2,2);
     \node[dot] (a) at (0,0) {};
     \node[dot] (b) at (1,0) {};
     \node[dot] (c) at (2,0) {};
     \node[dot] (a) at (0,1) {};
     \node[dot] (b) at (1,1) {};
     \node[dot] (c) at (2,1) {};
     \node[dot] (a) at (0,2) {};
     \node[dot] (b) at (1,2) {};
     \node[dot] (c) at (2,2) {};
 \draw (0.2,0.5) node[right]{{\tiny $\theta-\gamma$}};
 \draw (1.4,0.5) node[right]{{\tiny $\theta$}};
 \draw (0.4,1.5) node[right]{{\tiny $\theta$}};
 \draw (1.2,1.5) node[right]{{\tiny $\theta+\gamma$}};
\draw (1.1,-0.5) node[below]{$b_{-}$};
 \end{scope}
 
 \begin{scope}[xshift=10cm]
 \draw[help lines] (0,0) grid (2,2);
     \node[dot] (a) at (0,0) {};
     \node[dot] (b) at (1,0) {};
     \node[dot] (c) at (2,0) {};
     \node[dot] (a) at (0,1) {};
     \node[dot] (b) at (1,1) {};
     \node[dot] (c) at (2,1) {};
     \node[dot] (a) at (0,2) {};
     \node[dot] (b) at (1,2) {};
     \node[dot] (c) at (2,2) {};
 \draw (0.2,0.5) node[right]{{\tiny $\theta+\gamma$}};
 \draw (1.4,0.5) node[right]{{\tiny $\theta$}};
 \draw (0.4,1.5) node[right]{{\tiny $\theta$}};
 \draw (1.2,1.5) node[right]{{\tiny $\theta+\gamma$}};
\draw (1.1,-0.5) node[below]{$c_{+}$};
 \end{scope}
 
 \begin{scope}[xshift=12.5cm]
 \draw[help lines] (0,0) grid (2,2);
     \node[dot] (a) at (0,0) {};
     \node[dot] (b) at (1,0) {};
     \node[dot] (c) at (2,0) {};
     \node[dot] (a) at (0,1) {};
     \node[dot] (b) at (1,1) {};
     \node[dot] (c) at (2,1) {};
     \node[dot] (a) at (0,2) {};
     \node[dot] (b) at (1,2) {};
     \node[dot] (c) at (2,2) {};
 \draw (0.2,0.5) node[right]{{\tiny $\theta-\gamma$}};
 \draw (1.4,0.5) node[right]{{\tiny $\theta$}};
 \draw (0.4,1.5) node[right]{{\tiny $\theta$}};
 \draw (1.2,1.5) node[right]{{\tiny $\theta-\gamma$}};
\draw (1.1,-0.5) node[below]{$c_{-}$};
 \end{scope}

\begin{scope}
\path (0,1) coordinate (LL);
\path (1,1) coordinate (CC);
\path (2,1) coordinate (RR);
\path (1,0) coordinate (DD);
\path (1,2) coordinate (UU);
\draw[postaction=decorate] (LL) -- (CC);
\draw[postaction=decorate] (CC) -- (RR);
\draw[postaction=decorate] (DD) -- (CC);
\draw[postaction=decorate] (CC) -- (UU);
\node[dot] (a) at (CC) {};
\end{scope}

\begin{scope}[xshift=2.5cm]
\path (0,1) coordinate (LL);
\path (1,1) coordinate (CC);
\path (2,1) coordinate (RR);
\path (1,0) coordinate (DD);
\path (1,2) coordinate (UU);
\draw[postaction=decorate] (CC) -- (LL);
\draw[postaction=decorate] (RR) -- (CC);
\draw[postaction=decorate] (CC) -- (DD);
\draw[postaction=decorate] (UU) -- (CC);
\node[dot] (a) at (CC) {};
\end{scope}

\begin{scope}[xshift=5cm]
\path (0,1) coordinate (LL);
\path (1,1) coordinate (CC);
\path (2,1) coordinate (RR);
\path (1,0) coordinate (DD);
\path (1,2) coordinate (UU);
\draw[postaction=decorate] (LL) -- (CC);
\draw[postaction=decorate] (CC) -- (RR);
\draw[postaction=decorate] (CC) -- (DD);
\draw[postaction=decorate] (UU) -- (CC);
\node[dot] (a) at (CC) {};
\end{scope}

\begin{scope}[xshift=7.5cm]
\path (0,1) coordinate (LL);
\path (1,1) coordinate (CC);
\path (2,1) coordinate (RR);
\path (1,0) coordinate (DD);
\path (1,2) coordinate (UU);
\draw[postaction=decorate] (CC) -- (LL);
\draw[postaction=decorate] (RR) -- (CC);
\draw[postaction=decorate] (DD) -- (CC);
\draw[postaction=decorate] (CC) -- (UU);
\node[dot] (a) at (CC) {};
\end{scope}

\begin{scope}[xshift=10cm]
\path (0,1) coordinate (LL);
\path (1,1) coordinate (CC);
\path (2,1) coordinate (RR);
\path (1,0) coordinate (DD);
\path (1,2) coordinate (UU);
\draw[postaction=decorate] (LL) -- (CC);
\draw[postaction=decorate] (RR) -- (CC);
\draw[postaction=decorate] (CC) -- (DD);
\draw[postaction=decorate] (CC) -- (UU);
\node[dot] (a) at (CC) {};
\end{scope}

\begin{scope}[xshift=12.5cm]
\path (0,1) coordinate (LL);
\path (1,1) coordinate (CC);
\path (2,1) coordinate (RR);
\path (1,0) coordinate (DD);
\path (1,2) coordinate (UU);
\draw[postaction=decorate] (CC) -- (LL);
\draw[postaction=decorate] (CC) -- (RR);
\draw[postaction=decorate] (DD) -- (CC);
\draw[postaction=decorate] (UU) -- (CC);
\node[dot] (a) at (CC) {};
\end{scope}
\end{tikzpicture}
\caption{Face-vertex configurations and the associated Boltzmann weight.}
\label{fig:faces}
\end{figure}
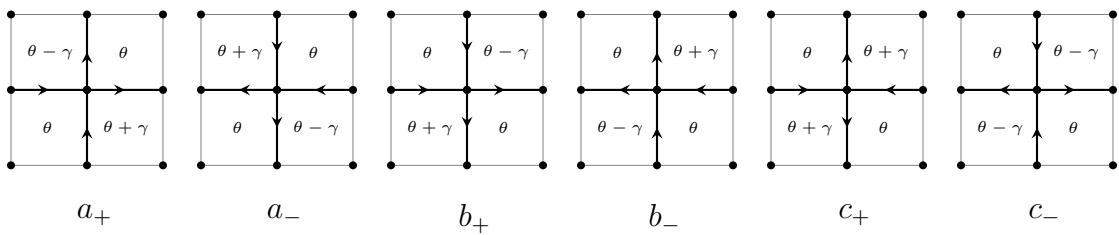

In what follows we will be dealing with a lattice formed by coloured faces  where each
adjacent face can not have the same colour. As it was remarked by Baxter in \cite{Baxter_1970}, this system 
can be thought as a system of particles interacting through an infinitely repulsive force between nearst neighbors
of the same type. For the trigonometric SOS model we will have $l_{i,j} = \theta + \bar{l}_{i,j} \gamma$, where $\bar{l}_{i,j}$
is an integer variable while $\theta$ and $\gamma$ are complex numbers. In the \figref{fig:faces} the height function $l_{i,j}$
characterising the colour of the face is projected into the center of the $(i,j)$ face.
Interestingly enough, it was remarked by Lenard \footnote{See {\it Note added in proof} of \cite{Lieb_1967}.} an equivalence between 
this colouring of faces and the configurations of a six vertex model. This equivalence is depicted in \figref{fig:faces} where basically
one removes the outer edges of the four-faces set and places arrows on the internal edges according to a certain rule. This rule is as follows: 
each face of a four-faces set is visited in the anticlockwise direction. If the colour changes by $+\gamma$ when intersecting
an edge, this edge receives an arrow pointing inwards; and if the colour changes by $-\gamma$ we place an arrow pointing
outwards on that edge. For a particular class of statistical weights this model is also called {\it Three-colouring model} 
\cite{Baxter_1970, Stroganov_1982}. 
In that case the height function $l_{i,j}$ is conveniently labeled by an element of the ring $\mathbb{Z}_3$. More precisely,
the $\mathbb{Z}_3$ structure labeling the colours of the faces is unveiled by considering the remainder after division of
$\bar{l}_{i,j}$ by $3$. 

\paragraph{The dynamical Yang-Baxter equation.} In \cite{Felder_94, Felder_1994, Felder_1996} it was demonstrated that the Boltzmann
weights of a face model satisfying (\ref{star_triang}) are encoded in the solutions of the dynamical 
Yang-Baxter equation. This dynamical version of the quantum Yang-Baxter equation was proposed by Felder in \cite{Felder_94} 
as the quantised form of a modified classical Yang-Baxter equation. In that case this modified classical Yang-Baxter
equation arises as the compatibility condition for the Knizhnik-Zamolodchikov-Bernard equations \cite{Bernard1_1988, Bernard2_1988}.
Previous to that, the dynamical Yang-Baxter equation had appeared in connection to the Liouville string field theory in \cite{Gervais_1984}.

Now let $\mathbb{V} = v_{+} \oplus v_{-}$ be a two-dimensional complex vector space and consider the operator
$\mathcal{R}(\lambda, \theta) \in \mbox{End} (\mathbb{V}  \otimes \mathbb{V} )$ with $\lambda, \theta \in \mathbb{C}$. The dynamical
Yang-Baxter equation for the trigonometric SOS model then reads
\<
\label{yb}
\mathcal{R}_{12}(\lambda_1 - \lambda_2, \theta - \gamma \hat{h}_3) \mathcal{R}_{13}(\lambda_1 - \lambda_3, \theta) \mathcal{R}_{23}(\lambda_2 - \lambda_3, \theta - \gamma \hat{h}_1 ) = \nonumber \\
\mathcal{R}_{23}(\lambda_2 - \lambda_3, \theta) \mathcal{R}_{13}(\lambda_1 - \lambda_3, \theta - \gamma \hat{h}_2) \mathcal{R}_{12}(\lambda_1 - \lambda_2,\theta) 
\>
where $\hat{h} = \mbox{diag}(1,-1)$. The Eq. (\ref{yb}) is defined in $\mbox{End} (\mathbb{V}_1  \otimes \mathbb{V}_2  \otimes \mathbb{V}_3 )$
and the action of $\mathcal{R}_{12}(\lambda, \theta - \gamma \hat{h}_3 )$ on $v_1 \otimes v_2 \otimes v_3$ is understood as
\[
\left[ \mathcal{R}(\lambda, \theta - \gamma h) v_1 \otimes v_2 \right] \otimes v_3 \; , 
\]
keeping in mind that $\hat{h}_3 v_3 = h v_3$. In other words, $h$ is simply the eigenvalue of $\hat{h}$ on the corresponding
subspace. The explicit trigonometric solution of (\ref{yb}) is given by
\[
\label{rmat}
\mathcal{R} (\lambda, \theta) = \left( \begin{matrix}
a_{+}(\lambda, \theta) & 0 & 0 & 0 \cr 
0 & b_{+}(\lambda, \theta) & c_{+}(\lambda, \theta) & 0 \cr
0 & c_{-}(\lambda, \theta) & b_{-}(\lambda, \theta) & 0 \cr
0 & 0 & 0 & a_{-}(\lambda, \theta) \end{matrix} \right)
\]
with non-null entries
\<
\label{bw}
a_{\pm}(\lambda, \theta) &=& \sinh{(\lambda + \gamma)} \nonumber \\
b_{\pm}(\lambda, \theta) &=& \sinh{(\lambda)} \frac{\sinh{(\theta \mp \gamma)}}{\sinh{(\theta)}} \nonumber  \\
c_{\pm}(\lambda, \theta) &=& \sinh{(\gamma)} \frac{\sinh{(\theta \mp \lambda)}}{\sinh{(\theta)}} \;\; .
\>
The solution described by (\ref{rmat}) and (\ref{bw}) consists of a particular limit of the elliptic solution
found in \cite{Felder_94, Felder_1994}. Also it is important to remark here that such solutions are in 
correspondence with Baxter's eight-vertex model after a vertex-face transformation \cite{Baxter_1971,Baxter_1973}.
Moreover, the dynamical $\mathcal{R}$-matrix (\ref{rmat}) satisfies the ice rule
\[
\label{ice}
\left[ \mathcal{R}_{ab}(\lambda, \theta), \hat{h}_{a} + \hat{h}_{b} \right] = 0 
\]
which plays an important role in estabilishing an algebra associated to dynamical $\mathcal{R}$-matrices.

Now we turn our attention to the relation between solutions of the dynamical Yang-Baxter equation (\ref{yb})
and the statistical weights of a SOS model satisfying (\ref{star_triang}). At this stage the observations of
Lenard \cite{Lieb_1967}, the vertex-face transformation introduced by Baxter \cite{Baxter_1973} and Felder's
dynamical Yang-Baxter equation \cite{Felder_1994} converge to the same point. Following  \cite{Felder_1994}
we thus have
\begin{align}
w_{\lambda} \left( \begin{matrix}   \theta - \gamma & \theta \cr \theta & \theta + \gamma \end{matrix} \right) & = a_{+} (\lambda, \theta) 
& w_{\lambda} \left( \begin{matrix}   \theta + \gamma & \theta \cr \theta & \theta - \gamma  \end{matrix} \right) & = a_{-} (\lambda, \theta) \nonumber \\
w_{\lambda} \left( \begin{matrix}   \theta - \gamma & \theta - 2\gamma \cr \theta & \theta - \gamma \end{matrix} \right) & = b_{+} (\lambda, \theta) 
& w_{\lambda} \left( \begin{matrix}   \theta + \gamma & \theta + 2\gamma \cr \theta & \theta + \gamma  \end{matrix} \right) & = b_{-} (\lambda, \theta) \nonumber \\
w_{\lambda} \left( \begin{matrix}   \theta - \gamma & \theta  \cr \theta & \theta - \gamma \end{matrix} \right) & = c_{+} (\lambda, \theta) 
& w_{\lambda} \left( \begin{matrix}   \theta + \gamma & \theta \cr \theta & \theta + \gamma  \end{matrix} \right) & = c_{-} (\lambda, \theta) \; \; .
\end{align}
This association is also depicted in \figref{fig:faces}.

\paragraph{The dynamical monodromy matrix.} Still following \cite{Felder_1994, Felder_1996} we define an inhomogeneous monodromy matrix $\mathcal{T}_{a}(\lambda, \theta)$ 
formed by the ordered product of dynamical $\mathcal{R}$-matrices. More precisely, this dynamical monodromy matrix reads
\[
\label{dmono}
\mathcal{T}_{a} (\lambda, \theta) = \prod_{i=1}^{L} \mathcal{R}_{a i}(\lambda - \mu_i, \theta_i) 
\]
with $\theta_i = \theta - \gamma \displaystyle \sum_{k=i+1}^L \hat{h}_{k}$, and it should not be confused with
the face monodromy matrix defined in (\ref{mono}).
 
The dynamical monodromy matrix (\ref{dmono}) is an operator living in the tensor product
space $\mathbb{V}_a \otimes \mathbb{V}_1 \otimes \dots \otimes \mathbb{V}_L$. The space $\mathbb{V}_a$ will be refereed to as auxiliar space
while the tensor product $\mathbb{V}_1 \otimes \dots \otimes \mathbb{V}_L$ will be called quantum space. In this way $\mathcal{T}_{a}(\lambda, \theta)$ can be regarded as
a matrix on the auxiliar space whose entries are matrices living in the quantum space. Here we shall restrict ourselves to the monodromy matrix built out of the
$\mathcal{R}$-matrix (\ref{rmat}). Therefore, it can be conveniently denoted as
\[
\label{abcd}
\mathcal{T}_{a} (\lambda, \theta) = \left( \begin{matrix}
A(\lambda, \theta) & B(\lambda, \theta) \cr
C(\lambda, \theta) & D(\lambda, \theta) \end{matrix} \right)  \; .
\]
Due to the dynamical Yang-Baxter equation (\ref{yb}) and the ice rule (\ref{ice}), one can
demonstrate that the monodromy matrix (\ref{dmono}) satisfies the algebraic relation
\<
\label{dyb}
\mathcal{R}_{ab}(\lambda_1 - \lambda_2, \theta - \gamma \gen{H}) \mathcal{T}_a (\lambda_1, \theta) \mathcal{T}_b (\lambda_2, \theta - \gamma \hat{h}_a) =  
\mathcal{T}_b (\lambda_2, \theta) \mathcal{T}_a (\lambda_1, \theta - \gamma \hat{h}_b ) \mathcal{R}_{ab}(\lambda_1 - \lambda_2, \theta) \nonumber \\
\>
where $\gen{H} = \displaystyle \sum_{k=1}^L \hat{h}_{k}$. With the help of the definition of $\hat{h}$, we can identify $\gen{H}$ 
as the Cartan element of the $\alg{su}(2)$ algebra on the tensor product space $\mathbb{V}_1 \otimes \dots \otimes \mathbb{V}_L$.
Furthermore, in the limit $\theta \rightarrow \infty$ we can immediately see that (\ref{rmat}) becomes the standard $\mathcal{R}$-matrix invariant under the 
quantum affine algebra $U_{q}[\widehat{\alg{su}}(2)]$. This observation extends to
the monodromy matrix (\ref{dmono}) and the algebra (\ref{dyb}), which become respectively the standard trigonometric six vertex model
monodromy matrix and Yang-Baxter algebra \cite{Sk_1979, Fad_1979}.

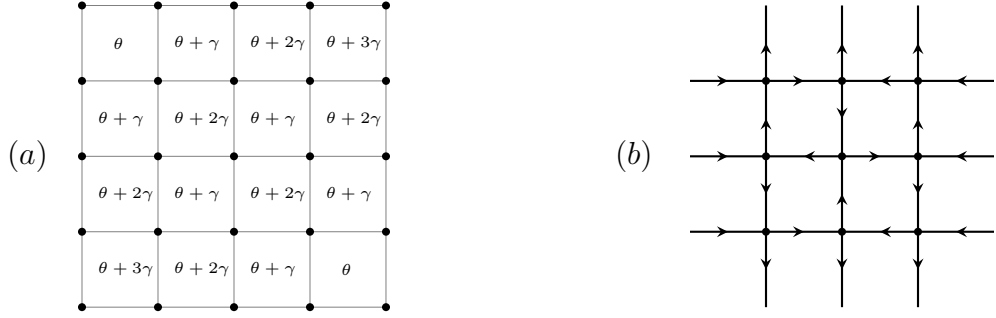
\begin{figure} \centering
\begin{tikzpicture}[inner sep=0pt,thick, dot/.style={fill=black,circle,minimum size=3pt}, >=stealth]
\draw[help lines] (0,0) grid (4,4);
    \node[dot] (a) at (0,0) {};
    \node[dot] (a) at (0,1) {};
    \node[dot] (a) at (0,2) {};
    \node[dot] (a) at (0,3) {};
    \node[dot] (a) at (0,4) {};
    \node[dot] (a) at (1,0) {};
    \node[dot] (a) at (1,1) {};
    \node[dot] (a) at (1,2) {};
    \node[dot] (a) at (1,3) {};
    \node[dot] (a) at (1,4) {};
    \node[dot] (a) at (2,0) {};
    \node[dot] (a) at (2,1) {};
    \node[dot] (a) at (2,2) {};
    \node[dot] (a) at (2,3) {};
    \node[dot] (a) at (2,4) {};
    \node[dot] (a) at (3,0) {};
    \node[dot] (a) at (3,1) {};
    \node[dot] (a) at (3,2) {};
    \node[dot] (a) at (3,3) {};
    \node[dot] (a) at (3,4) {};
    \node[dot] (a) at (4,0) {};
    \node[dot] (a) at (4,1) {};
    \node[dot] (a) at (4,2) {};
    \node[dot] (a) at (4,3) {};
    \node[dot] (a) at (4,4) {};

\draw (0.4,3.5) node[right]{{\tiny $\theta$}};
\draw (0.2,2.5) node[right]{{\tiny $\theta+\gamma$}};
\draw (0.2,1.5) node[right]{{\tiny $\theta+2\gamma$}};
\draw (0.2,0.5) node[right]{{\tiny $\theta+3\gamma$}};

\draw (1.2,3.5) node[right]{{\tiny $\theta+\gamma$}};
\draw (1.2,2.5) node[right]{{\tiny $\theta+2\gamma$}};
\draw (1.2,1.5) node[right]{{\tiny $\theta+\gamma$}};
\draw (1.2,0.5) node[right]{{\tiny $\theta+2\gamma$}};

\draw (2.2,3.5) node[right]{{\tiny $\theta+2\gamma$}};
\draw (2.2,2.5) node[right]{{\tiny $\theta+\gamma$}};
\draw (2.2,1.5) node[right]{{\tiny $\theta+2\gamma$}};
\draw (2.2,0.5) node[right]{{\tiny $\theta+\gamma$}};

\draw (3.2,3.5) node[right]{{\tiny $\theta+3\gamma$}};
\draw (3.2,2.5) node[right]{{\tiny $\theta+2\gamma$}};
\draw (3.2,1.5) node[right]{{\tiny $\theta+\gamma$}};
\draw (3.4,0.5) node[right]{{\tiny $\theta$}};

\draw (-1,2) node[right]{$(a)$} ;

\begin{scope}[xshift=8cm]
\foreach \i in {0,...,4}
{
    \foreach \j in {0,...,4}
  {
      \path (\i,\j) coordinate (P\i\j);
  }
}
\begin{scope}[decoration={
    markings,
    mark=at position 0.5 with {\arrow{>}}}
    ]
    \draw[postaction={decorate}] (P01)--(P11);
    \draw[postaction={decorate}] (P11)--(P21);
    \draw[postaction={decorate}] (P31)--(P21);
    \draw[postaction={decorate}] (P41)--(P31);
    \draw[postaction={decorate}] (P02)--(P12);
    \draw[postaction={decorate}] (P22)--(P12);
    \draw[postaction={decorate}] (P22)--(P32);
    \draw[postaction={decorate}] (P42)--(P32);
    \draw[postaction={decorate}] (P03)--(P13);
    \draw[postaction={decorate}] (P13)--(P23);
    \draw[postaction={decorate}] (P33)--(P23);
    \draw[postaction={decorate}] (P43)--(P33);
    \draw[postaction={decorate}] (P11)--(P10);
    \draw[postaction={decorate}] (P12)--(P11);
    \draw[postaction={decorate}] (P12)--(P13);
    \draw[postaction={decorate}] (P13)--(P14);
    \draw[postaction={decorate}] (P21)--(P20);
    \draw[postaction={decorate}] (P21)--(P22);
    \draw[postaction={decorate}] (P23)--(P22);
    \draw[postaction={decorate}] (P23)--(P24);
    \draw[postaction={decorate}] (P31)--(P30);
    \draw[postaction={decorate}] (P32)--(P31);
    \draw[postaction={decorate}] (P32)--(P33);
    \draw[postaction={decorate}] (P33)--(P34);

    \node[dot] (a) at (P11) {};
    \node[dot] (a) at (P12) {};
    \node[dot] (a) at (P13) {};
    \node[dot] (a) at (P21) {};
    \node[dot] (a) at (P22) {};
    \node[dot] (a) at (P23) {};
    \node[dot] (a) at (P31) {};
    \node[dot] (a) at (P32) {};
    \node[dot] (a) at (P33) {};

\draw (-1,2) node[right]{$(b)$} ;
\end{scope}
\end{scope}
\end{tikzpicture}
\caption{$(a)$ SOS model with domain wall boundaries. $(b)$ The corresponding structure of vertices.}
\label{fig:sos_dw}
\end{figure}

\paragraph{Domain wall boundary conditions.} We shall now consider the trigonometric SOS model 
on the lattice described in \figref{fig:lattice} with $N=M=L+1$ and special boundary conditions. As for the boundary conditions,
we set $l_{1,j} = l_{j,1} = \theta + (L+1-j)\gamma$ and $l_{L+1,j} = l_{j,L+1} = \theta + (j-1)\gamma$. This special boundary
conditions is illustrated in \figref{fig:sos_dw} together with the corresponding structure of vertices.
In the vertices language we can immediately
recognize this special boundary conditions as the case of domain wall boundaries introduced by Korepin in \cite{Korepin82}. This observation
if of fundamental importance allowing us to express the model partition function in terms of the entries of the monodromy matrix (\ref{abcd})
analogously to the case of the standard six vertex model \cite{Korepin82}. The diagrammatical interpretation of (\ref{dmono}) is given
in \cite{Felder_1994, Felder_1996} and for a discussion on the construction of the partition function (\ref{pf}) in terms of the
components (\ref{abcd}) we refer to \cite{deGier_Galleas11}. For the trigonometric SOS model considered here we only have
to keep in mind that the entries of the dynamical monodromy matrix (\ref{abcd}) also depends on the colour variable 
$\theta$ governed by the height functions $l_{j,1}$. In this way the partition function (\ref{pf}) for the trigonometric
SOS model with domain wall boundaries can be written as
\[
\label{pft}
Z_{\theta} = \bra{\bar{0}} \prod_{j=1}^{L} B(\lambda_j, \theta + j \gamma) \ket{0} 
\]
where
\[
\label{states1}
\ket{0} = \bigotimes_{i=1}^{L} \left( \begin{matrix}
1 \cr
0 \end{matrix} \right) \qquad \qquad \mbox{and} \qquad \qquad
\ket{\bar{0}} = \bigotimes_{i=1}^{L} \left( \begin{matrix}
0 \cr
1 \end{matrix} \right) \; .
\]

\section{Dynamical Yang-Baxter algebra and functional relations}
\label{sec:dyba}

The dynamical Yang-Baxter algebra (\ref{dyb}) encodes commutation rules for the entries of the dynamical monodromy matrix (\ref{dmono}). 
In contrast to the standard Yang-Baxter algebra, the relation (\ref{dyb}) not only contains the generators $A(\lambda, \theta)$, $B(\lambda, \theta)$,
$C(\lambda, \theta)$ and $D(\lambda, \theta)$, but also the $\alg{su}(2)$ Cartan generator $\gen{H}$.
This indicates that the algebra defined by (\ref{dyb}) needs to be complemented. 

\paragraph{The $\alg{su}(2)$ Cartan generator.} The definition of $\gen{H}$ together with (\ref{dmono}) and (\ref{abcd}) allow us to directly compute the commutators 
between the $\alg{su}(2)$ Cartan generator $\gen{H}$ and the entries of the monodromy matrix (\ref{abcd}). In the limit $\theta \rightarrow \infty$ this analysis
has been performed in \cite{Fad_1979,Korepin82} and here we find that there are no significant modifications for arbitrary $\theta$. 
Nevertheless, this analysis can also be found in \appref{sec:high} and we have
\begin{align}
\label{ybsu}
\left[ A(\lambda, \theta) , \gen{H} \right] &= 0 &  \left[ B(\lambda, \theta) , \gen{H} \right] &= 2  B(\lambda, \theta) \nonumber \\
\left[ C(\lambda, \theta) , \gen{H} \right] &= -2 C(\lambda, \theta)&  \left[ D(\lambda, \theta) , \gen{H} \right] &= 0 \; .
\end{align}
From the $\mathcal{R}$-matrix (\ref{rmat}) we can see that the generator $\gen{H}$ will appear in (\ref{dyb}) only as $e^{\pm \gamma \gen{H}}$. 
Thus for convenience we define the operator $\gen{K} = q^{\gen{H}}$ with $q = e^{\gamma}$, in such a way that the commutation rules
(\ref{ybsu}) imply the following relations:
\begin{align}
\label{ybsuk}
B(\lambda, \theta) \gen{K}  &= q^2  \gen{K} B(\lambda, \theta) & \left[ A(\lambda, \theta) , \gen{K} \right] &= 0 \nonumber \\
C(\lambda, \theta) \gen{K}  &= q^{-2}  \gen{K} C(\lambda, \theta) &  \left[ D(\lambda, \theta) , \gen{K} \right] &= 0 \; .
\end{align}

\medskip

The commutation rules (\ref{ybsuk}) in addition to (\ref{dyb}) form an extended Yang-Baxter algebra with generators $A(\lambda, \theta)$, $B(\lambda, \theta)$,
$C(\lambda, \theta)$, $D(\lambda, \theta)$ and $\gen{K}^{\pm}$. For our purposes here we only need to extract a few commutation relations from (\ref{dyb}).
Namely,
\<
B(\lambda_1, \theta) B(\lambda_2, \theta +\gamma) &=& B(\lambda_2, \theta) B(\lambda_1, \theta +\gamma) \nonumber \\
A(\lambda_1, \theta + \gamma) B(\lambda_2, \theta) &=& \frac{s(\lambda_2 - \lambda_1 +\gamma)}{s(\lambda_2 - \lambda_1)} \frac{s(\theta +\gamma)}{s(\theta + 2 \gamma)} B(\lambda_2, \theta + \gamma) A(\lambda_1, \theta + 2\gamma) \nonumber \\
&-& \frac{s(\theta + \gamma - \lambda_2 + \lambda_1)}{s(\lambda_2 - \lambda_1)} \frac{s(\gamma)}{s(\theta + 2\gamma)} B(\lambda_1, \theta + \gamma) A(\lambda_2, \theta + 2\gamma) \nonumber 
\>
\<
&& D(\lambda_1, \theta - \gamma) B(\lambda_2, \theta) = \nonumber \\
&& \frac{s(\lambda_1 - \lambda_2 +\gamma)}{s(\lambda_1 - \lambda_2)} B(\lambda_2, \theta-\gamma) D(\lambda_1, \theta) [t q \gen{K}^{-1} - t^{-1} q^{-1} \gen{K} ] 
[t q^2 \gen{K}^{-1} - t^{-1} q^{-2} \gen{K} ]^{-1} \nonumber \\
&& - \frac{s(\gamma)}{s(\lambda_1 - \lambda_2)} B(\lambda_1, \theta-\gamma) D(\lambda_2, \theta) [ t q \bar{x}_1 \bar{x}_2^{-1} \gen{K}^{-1} - t^{-1} q^{-1} \bar{x}_1^{-1} \bar{x}_2 \gen{K} ] [t q^2 \gen{K}^{-1} - t^{-1} q^{-2} \gen{K} ]^{-1} \nonumber \\
\nonumber \\
&& C(\lambda_1, \theta + \gamma) B(\lambda_2, \theta) = \nonumber \\
&& \frac{s(\theta)}{s(\theta + \gamma)}  B(\lambda_2, \theta + \gamma) C(\lambda_1, \theta + 2\gamma)[t q \gen{K}^{-1} - t^{-1} q^{-1} \gen{K} ] [t q^2 \gen{K}^{-1} - t^{-1} q^{-2} \gen{K} ]^{-1} \nonumber \\
&&+ \frac{s(\gamma)}{s(\theta + \gamma)} \frac{s(\theta + \gamma + \lambda_1 - \lambda_2)}{s(\lambda_1 - \lambda_2)} A(\lambda_2, \theta + \gamma) D(\lambda_1, \theta) [t q \gen{K}^{-1} - t^{-1} q^{-1} \gen{K} ] [t q^2 \gen{K}^{-1} - t^{-1} q^{-2} \gen{K} ]^{-1} \nonumber \\
&&- \frac{s(\gamma)}{s(\lambda_1 - \lambda_2)} A(\lambda_1, \theta + \gamma) D(\lambda_2, \theta) [ t q \bar{x}_1 \bar{x}_2^{-1} \gen{K}^{-1} - t^{-1} q^{-1} \bar{x}_1^{-1} \bar{x}_2 \gen{K} ] [t q^2 \gen{K}^{-1} - t^{-1} q^{-2} \gen{K} ]^{-1} \; . \nonumber \\
\label{commut}
\>
In the above commutation rules we have employed the notation $s(\lambda) = \sinh(\lambda)$, $t = e^{\theta}$ and $\bar{x}_i = e^{\lambda_i}$. 
The algebra formed by the relations (\ref{ybsuk}) and (\ref{commut}) will be one of the main ingredients in the derivation of a functional relation determining the 
partition function (\ref{pft}). 

In order to proceed we will also need to consider the action of the generators $A(\lambda, \theta)$, $B(\lambda, \theta)$, $C(\lambda, \theta)$, $D(\lambda, \theta)$ and $\gen{K}$ on the states
$\ket{0}$ and $\ket{\bar{0}}$ defined in (\ref{states1}). Those states are the $\alg{su}(2)$ highest and lowest weight states respectively, and from (\ref{rmat}), (\ref{dmono}) and (\ref{abcd}) we readly obtain the relations
\<
\label{action}
\gen{K}^{\pm} \ket{0} &=& q^{\pm L} \ket{0} \nonumber \\
A(\lambda, \theta) \ket{0} &=& \prod_{i=1}^{L} s(\lambda - \mu_i + \gamma) \ket{0} \nonumber \\
D(\lambda, \theta) \ket{0} &=& \frac{s(\theta + \gamma)}{s(\theta - (L-1)\gamma)} \prod_{i=1}^{L} s(\lambda - \mu_i) \ket{0} \; . \nonumber \\
\>
Moreover, we also obtain the properties 
\begin{align}
\label{destr}
B(\lambda, \theta) \ket{0} &= \dagger & C(\lambda, \theta) \ket{0} &= 0 \nonumber \\
B(\lambda, \theta) \ket{\bar{0}} &= 0 & C(\lambda, \theta) \ket{\bar{0}} &= \ddagger \; ,
\end{align}
where the symbols $\dagger$ and $\ddagger$ stand for non-null values. The relations (\ref{destr})
together with (\ref{ybsu}) support considering $B(\lambda, \theta)$ and $C(\lambda, \theta)$ as creation and annihilation 
operators respectively with respect to the pseudo-vacuum state $\ket{0}$.

Altogether the relations (\ref{ybsuk})-(\ref{destr}) allow us to derive the following formula,
\<
\label{cbb}
C(\lambda_0, \theta + \gamma) && \prod_{i=1}^{n} B(\lambda_i , \theta + (i-1)\gamma ) \ket{0} = 
\sum_{i=1}^{n} M_{i} \prod_{j=1}^{n-1} B(\lambda_{r_j^{(i)}}, \theta + j \gamma) \ket{0} \nonumber \\
&+& \sum_{j=2}^{n} \sum_{i=1}^{j-1} N_{ji} B(\lambda_0 , \theta + \gamma) \prod_{k=1}^{n-2}
B(\lambda_{s_k^{(ij)}}, \theta + (k+1) \gamma) \ket{0} \; , \nonumber \\
\>
where 
\[
r_j^{(i)} = \begin{cases}
j \qquad \qquad 1 \leq j < i \cr
j+1  \qquad \; i \leq j \leq n-1 \end{cases} 
\qquad \mbox{and}
\qquad 
s_k^{(ij)} = \begin{cases}
k \qquad \qquad 1 \leq k < i \cr
k+1  \qquad \; i \leq k < j \cr
k+2  \qquad \; j \leq k \leq n-2 
\end{cases} \; .
\]
In their turn the coefficients $M_{i}$ and $N_{ji}$ are given by
\<
\label{mm}
&& M_i = \nonumber \\
&& \frac{s(\gamma)}{s(\lambda_i - \lambda_0)} \frac{s(\theta + \gamma)}{s(\theta + n \gamma)}\frac{s(\lambda_0 - \lambda_i + \theta + (2n-1 -L)\gamma )}{s(\theta + (2n-1-L)\gamma )} \frac{s(\theta +(n-L)\gamma)}{s(\theta + (2n-L)\gamma)} \frac{s(\theta + n\gamma)}{s(\theta + (n-L)\gamma)} \nonumber \\
&& \times \prod_{l=1}^{L} s(\lambda_0 - \mu_l + \gamma) s(\lambda_i - \mu_l) \prod_{\stackrel{k=1}{k\neq i}}^{n} \frac{s(\lambda_i - \lambda_k + \gamma)}{s(\lambda_i - \lambda_k)} \frac{s(\lambda_k - \lambda_0 + \gamma)}{s(\lambda_k - \lambda_0)} \nonumber \\
&& + \frac{s(\gamma)}{s(\lambda_0 - \lambda_i)} \frac{s(\lambda_0 - \lambda_i + \theta + \gamma)}{s(\theta + n \gamma)}\frac{s(\theta + (n-L)\gamma )}{s(\theta + (2n-L)\gamma )}  
\frac{s(\theta + n\gamma)}{s(\theta + (n-L)\gamma)} \nonumber \\
&& \times \prod_{l=1}^{L} s(\lambda_i - \mu_l + \gamma) s(\lambda_0 - \mu_l) \prod_{\stackrel{k=1}{k\neq i}}^{n} \frac{s(\lambda_0 - \lambda_k + \gamma)}{s(\lambda_0 - \lambda_k)} \frac{s(\lambda_k - \lambda_i + \gamma)}{s(\lambda_k - \lambda_i)}  \nonumber \\
\>
\<
\label{nn}
&& N_{ji} = \nonumber \\
&& \frac{s(\gamma)}{s(\lambda_0 - \lambda_j)} \frac{s(\gamma)}{s(\lambda_i - \lambda_0)} \frac{s(\lambda_j - \lambda_i + \gamma)}{s(\lambda_j - \lambda_i)}
\frac{s(\lambda_0 - \lambda_i + \theta + \gamma)}{s(\theta + n \gamma)} \frac{s(\lambda_0 - \lambda_j + \theta + (2n-1-L)\gamma)}{s(\theta + (2n-1-L)\gamma)} \nonumber \\
&& \times \frac{s(\theta + (n-L)\gamma)}{s(\theta + (2n-L)\gamma)} \frac{s(\theta + n\gamma)}{s(\theta + (n-L)\gamma)} \prod_{l=1}^{L} s(\lambda_i - \mu_l + \gamma) s(\lambda_j - \mu_l) \nonumber \\
&& \times \prod_{\stackrel{m=1}{m\neq i,j}}^{n} \frac{s(\lambda_j - \lambda_m + \gamma)}{s(\lambda_j - \lambda_m)} \frac{s(\lambda_m - \lambda_i + \gamma)}{s(\lambda_m - \lambda_i)}  \nonumber \\
&& + \frac{s(\gamma)}{s(\lambda_0 - \lambda_i)} \frac{s(\gamma)}{s(\lambda_j - \lambda_0)} \frac{s(\lambda_i - \lambda_j + \gamma)}{s(\lambda_i - \lambda_j)}
\frac{s(\lambda_0 - \lambda_j + \theta + \gamma)}{s(\theta + n \gamma)} \frac{s(\lambda_0 - \lambda_i + \theta + (2n-1-L)\gamma)}{s(\theta + (2n-1-L)\gamma)} \nonumber \\
&& \times \frac{s(\theta + (n-L)\gamma)}{s(\theta + (2n-L)\gamma)} \frac{s(\theta + n\gamma)}{s(\theta + (n-L)\gamma)} \prod_{l=1}^{L} s(\lambda_j - \mu_l + \gamma) s(\lambda_i - \mu_l) \nonumber \\
&& \times \prod_{\stackrel{m=1}{m\neq i,j}}^{n} \frac{s(\lambda_i - \lambda_m + \gamma)}{s(\lambda_i - \lambda_m)} \frac{s(\lambda_m - \lambda_j + \gamma)}{s(\lambda_m - \lambda_j)} \; . \nonumber \\
\>
As expected, the expressions (\ref{cbb})-(\ref{nn}) recover the ones presented in \cite{Korepin82,Galleas10} in the limit $\theta \rightarrow \infty$.
At this stage we have gathered most of the ingredients required to obtain a functional equation determining the partition function (\ref{pft}).

\paragraph{Functional Equation.} The relation (\ref{cbb}) is valid for the product of an arbitrary number of operators $B(\lambda_i, \theta)$,
and following the approach devised in \cite{Galleas10}, we examine the quantity 
\[
\bra{\bar{0}} C(\lambda_0, \theta + \gamma) \prod_{i=1}^{L+1} B(\lambda_i , \theta + (i-1)\gamma ) \ket{0}
\]
under the light of the extended Yang-Baxter algebra formed by (\ref{dyb}) and (\ref{ybsuk}). Thus considering (\ref{cbb})
and the definition (\ref{pft}) we immediately obtain the relation
\<
\label{fe0}
\bra{\bar{0}} C(\lambda_0, \theta + \gamma) && \prod_{i=1}^{L+1} B(\lambda_i , \theta + (i-1)\gamma ) \ket{0} = \sum_{i}^{L+1} M_i Z_{\theta}(\lambda_1, \dots , \lambda_{i-1} , \lambda_{i+1} , \dots , \lambda_{L+1}) \nonumber \\
&+& \sum_{j=2}^{L+1} \sum_{i=1}^{j-1} N_{ji} Z_{\theta}(\lambda_0, \dots , \lambda_{i-1} , \lambda_{i+1} , \dots , \lambda_{j-1} , \lambda_{j+1} ,\dots , \lambda_{L+1}) \; ,
\>
where the coefficients $M_i$ and $N_{ji}$ are the ones given by (\ref{mm}) and (\ref{nn}) with $n=L+1$. 
On the other hand, the highest weight representation theory of the $\alg{su}(2)$ algebra tells us that the LHS of (\ref{fe0}) vanishes. This property
has been already discussed in \cite{Galleas10} but for completeness we also include it in the \appref{sec:high}.
In this way we are left with the following functional equation for the partition function (\ref{pft}),
\<
\label{FZ}
&&\sum_{i=1}^{L+1} M_i Z_{\theta}(\lambda_1, \dots , \lambda_{i-1} , \lambda_{i+1} , \dots , \lambda_{L+1}) \nonumber \\
&& \qquad \qquad  + \sum_{j=2}^{L+1} \sum_{i=1}^{j-1} N_{ji} Z_{\theta}(\lambda_0, \dots , \lambda_{i-1} , \lambda_{i+1} , \dots , \lambda_{j-1} , \lambda_{j+1} ,\dots , \lambda_{L+1}) = 0 \; . \nonumber \\
\>
Some comments are in order at this stage. The partition function $Z_{\theta}$ depends on two sets of variables, $\{ \lambda_i \}$ and $\{ \mu_i \}$, as well as parameters
$\gamma$ and $\theta$. Nevertheless, within this approach we can see that the set of variables $\{ \mu_i \}$ can also be regarded as parameters. 
The Eq. (\ref{FZ}) is linear and homogeneous in $Z_{\theta}$ and this observation will have important consequences for the characterisation of the desired solution. 
The homogeneity of (\ref{FZ}) tells us that if $Z_{\theta}$ is a solution, so is $\alpha Z_{\theta}$ where $\alpha$ is a constant. In fact $\alpha$ only needs to be
independent of the variables $\{ \lambda_i \}$. Therefore, the Eq. (\ref{FZ}) can determine the partition function at most up to a constant and the full determination of the partition function will require that we are at least able to compute $Z_{\theta}$ for some particular value of the variables $\lambda_i$. Moreover, the linearity of (\ref{FZ}) tells us that if $Z_{\theta}^{(1)}$ and $Z_{\theta}^{(2)}$ are two solutions of (\ref{FZ}), the linear combination $Z_{\theta}^{(1)} + Z_{\theta}^{(2)}$ is also a solution. This fact suggests that classifying the classes of unique solutions of (\ref{FZ}) is an important step in this framework. The Eq. (\ref{FZ}) has the same structure of the functional equation derived in \cite{Galleas10}, however the coefficients $M_i$ and $N_{ji}$ are deformed by the dynamical parameter $\theta$. 
In what follows these issues will be discussed and the desired solution of (\ref{FZ}) will be presented.

\section{The partition function}
\label{sec:partfun}

In this section we shall consider extra properties expected for the partition function (\ref{pft}) which will enable us to 
select the appropriate solution of (\ref{FZ}). These properties include the multivariate polynomial structure of the
partition function $Z_{\theta}$, as well as its asymptotic behaviour. In order to avoid an overcrowded section we shall discuss
the mentioned properties in the \appref{sec:polstruc} and here we present only the required results. 

\paragraph{Polynomial structure.}  The partition function $Z_{\theta}$ is of the form,
\[
\label{pol}
Z_{\theta}(\lambda_1 , \dots , \lambda_L) = \frac{\bar{Z}_{\theta} (x_1 , \dots , x_L)}{\displaystyle \prod_{i=1}^{L} x^{\frac{L}{2}}_i} \; ,
\]
where $\bar{Z}_{\theta}(x_1 , \dots , x_L)$ is a polynomial of order $L$ in each variable $x_i = e^{2\lambda_i}$ separately. The same 
polynomial structure holds if we consider the variables $\{ \mu_i \}$ instead of $\{ \lambda_i \}$, although this property will
not be required.

\paragraph{Asymptotic behaviour.} In the limit $x_i \rightarrow \infty$, the function $\bar{Z}_{\theta}$ possesses the
asymptotic behaviour
\[
\label{asym}
\bar{Z} \sim \frac{(q-q^{-1})^L}{2^{L^2}} \frac{[ L ]_{q^2} ! }{\displaystyle \prod_{n=1}^{L} (1 - q^{2n} t^2 ) u_{n}^{\frac{L}{2}}}  (x_1 \dots x_L)^L  \; ,
\]
where $[ n ]_q ! = 1 (1+q)(1+q+q^2) \dots (1+q+ \dots +q^n)$ denotes the $q$-factorial function and $u_i = e^{2 \mu_i}$. 

\medskip

For the moment we leave the properties (\ref{pol}) and (\ref{asym}) at rest and proceed with a more careful examination of the
functional equation (\ref{FZ}). Firstly we notice that besides the set of variables $\{ \lambda_1, \lambda_2, \dots , \lambda_L \}$ required
to characterise its solution, the Eq. (\ref{FZ}) 
also depends on variables $\lambda_0$ and $\lambda_{L+1}$. 
Thus the variables $\lambda_0$ and $\lambda_{L+1}$ can be fixed in order to fulfill our needs.
In particular, those variables can be chosen
in such a way that solving (\ref{FZ}) under the conditions (\ref{pol}) and (\ref{asym}) becomes systematic and simple. For illustrative purposes,
let us see how this approach would work in the case $L=2$.
In that case our functional equation reads
\<
\label{L2}
&& M_1 Z_{\theta}(\lambda_2,\lambda_3) + M_2 Z_{\theta}(\lambda_1,\lambda_3) + M_3 Z_{\theta}(\lambda_1,\lambda_2) \nonumber \\
&& + N_{21} Z_{\theta}(\lambda_0,\lambda_3) + N_{31} Z_{\theta}(\lambda_0,\lambda_2) + N_{32} Z_{\theta}(\lambda_0,\lambda_1) = 0 
\>
and we set $\lambda_0 = \mu_1$ and $\lambda_3 = \mu_1 - \gamma$. By doing so we find 
\[
\left. M_1 \right|_{\stackrel{\lambda_0 = \mu_1}{\lambda_3 = \mu_1 - \gamma}} = \left. M_2 \right|_{\stackrel{\lambda_0 = \mu_1}{\lambda_3 = \mu_1 - \gamma}} = 0
\]
and we also define
\<
m_2 = \left. M_3 \right|_{\stackrel{\lambda_0 = \mu_1}{\lambda_3 = \mu_1 - \gamma}} &=& - \frac{s(\theta + \gamma)}{s(\theta + 3\gamma)} s(\gamma)^2 s(\mu_1 - \mu_2 + \gamma) s(\mu_2 - \mu_1 + \gamma) \nonumber \\
\bar{m}_1 = \left. N_{31} \right|_{\stackrel{\lambda_0 = \mu_1}{\lambda_3 = \mu_1 - \gamma}} &=& \frac{s(\theta + \gamma - \lambda_1 + \mu_1)}{s(\theta + 3\gamma)}  s(\gamma)^2  s(\mu_2 - \mu_1 + \gamma) s(\lambda_1 -\mu_2 + \gamma) \nonumber \\
&& \times \frac{s(\lambda_2 - \mu_1)}{s(\lambda_2 - \mu_1+\gamma)} \frac{s(\lambda_2 - \lambda_1 + \gamma)}{s(\lambda_2 - \lambda_1)} \nonumber \\
\bar{m}_2 = \left. N_{32} \right|_{\stackrel{\lambda_0 = \mu_1}{\lambda_3 = \mu_1 - \gamma}} &=& \frac{s(\theta + \gamma - \lambda_2 + \mu_1)}{s(\theta + 3\gamma)}  s(\gamma)^2  s(\mu_2 - \mu_1 + \gamma) s(\lambda_2 -\mu_2 + \gamma) \nonumber \\
&& \times \frac{s(\lambda_1 - \mu_1)}{s(\lambda_1 - \mu_1+\gamma)} \frac{s(\lambda_1 - \lambda_2 + \gamma)}{s(\lambda_1 - \lambda_2)}\; .
\>
With the above specialisation of the variables $\lambda_0$ and $\lambda_3$, the Eq. (\ref{L2}) reduces to
\<
\label{quase}
Z_{\theta}(\lambda_1 , \lambda_2) &=& \frac{s(\theta + \gamma -\lambda_2 + \mu_1)}{s(\theta+\gamma)} \frac{s(\lambda_2 - \mu_2 + \gamma)}{s(\mu_1 - \mu_2 + \gamma)}
\frac{s(\lambda_1 - \mu_1)}{s(\lambda_1 - \mu_1 + \gamma)} \frac{s(\lambda_1 - \lambda_2+\gamma)}{s(\lambda_1 - \lambda_2)} Z_{\theta}(\mu_1 , \lambda_1) \nonumber \\
&+& \frac{s(\theta + \gamma -\lambda_1 + \mu_1)}{s(\theta+\gamma)} \frac{s(\lambda_1 - \mu_2 + \gamma)}{s(\mu_1 - \mu_2 + \gamma)}
\frac{s(\lambda_2 - \mu_1)}{s(\lambda_2 - \mu_1 + \gamma)} \frac{s(\lambda_2 - \lambda_1+\gamma)}{s(\lambda_2 - \lambda_1)} Z_{\theta}(\mu_1 , \lambda_2) \nonumber \\
&-& \left[ \frac{N_{21}}{M_3} \right]_{\stackrel{\lambda_0 = \mu_1}{\lambda_3 = \mu_1 - \gamma}} Z_{\theta}(\mu_1 , \mu_1 - \gamma) \; .
\>
In addition to that the Eq. (\ref{quase}) reduces to the following identity when $\lambda_2 = \mu_1$,
\[
\label{zz}
\left[ \frac{N_{21}}{M_3} \right]_{\stackrel{\lambda_0, \lambda_2 = \mu_1}{\lambda_3 = \mu_1 - \gamma}} Z_{\theta}(\mu_1 , \mu_1 - \gamma) = Z_{\theta}(\mu_1, \lambda_1) - Z_{\theta}(\lambda_1, \mu_1) \; .
\]
As demonstrated in the \appref{sec:symmetries}, the functional equation (\ref{FZ}) admits only symmetric solutions, i.e. $Z_{\theta}(\lambda_1 , \lambda_2) = Z_{\theta}(\lambda_2 , \lambda_1)$, and consequently the RHS of (\ref{zz}) vanishes. As the quantity $\left[ \frac{N_{21}}{M_3} \right]_{\stackrel{\lambda_0 , \lambda_2 = \mu_1}{\lambda_3 = \mu_1 - \gamma}}$ is finite we can thus conclude that $Z_{\theta}(\mu_1 , \mu_1 - \gamma) = 0$. This property simplifies (\ref{quase}) and we are left with
\begin{align}
\label{quase1}
& s(\mu_1 - \mu_2 + \gamma) s(\theta+\gamma) Z_{\theta}(\lambda_1 , \lambda_2) = \nonumber \\
& s(\theta + \gamma -\lambda_2 + \mu_1) s(\lambda_2 - \mu_2 + \gamma) \frac{s(\lambda_1 - \mu_1)}{s(\lambda_1 - \mu_1 + \gamma)} \frac{s(\lambda_1 - \lambda_2+\gamma)}{s(\lambda_1 - \lambda_2)} Z_{\theta}(\mu_1 , \lambda_1) \; + \nonumber \\
& s(\theta + \gamma -\lambda_1 + \mu_1) s(\lambda_1 - \mu_2 + \gamma) \frac{s(\lambda_2 - \mu_1)}{s(\lambda_2 - \mu_1 + \gamma)} \frac{s(\lambda_2 - \lambda_1+\gamma)}{s(\lambda_2 - \lambda_1)} Z_{\theta}(\mu_1 , \lambda_2) \; . \nonumber \\
\end{align}

The vanishing condition of $Z_{\theta}$ above unveiled have a special appeal since we are interested in the polynomial solution of (\ref{L2}) with order dictated by (\ref{pol}). 
For instance, they allow us to write
\[
\label{p2}
\left. Z_{\theta}(\lambda_1, \lambda_2 ) \right|_{\lambda_1 = \mu_1} = s(\lambda_2 - \mu_1 + \gamma)  V(\lambda_2 ) 
\] 
where now $V(\lambda_2 )$ needs to be a polynomial of the same order as $Z_{\theta}$ with $L=1$ in order to satisfy (\ref{pol}).
The expression (\ref{p2}) can now be replaced in (\ref{quase1}) yielding the following relation,
\begin{align}
\label{quase2}
& Z_{\theta}(\lambda_1 , \lambda_2 | \mu_1 , \mu_2) = \nonumber \\
&  \frac{s(\theta + \gamma -\lambda_2 + \mu_1)}{s(\theta+\gamma)} \frac{s(\lambda_2 - \mu_2 + \gamma)}{s(\mu_1 - \mu_2 + \gamma)} s(\lambda_1 - \mu_1) \frac{s(\lambda_1 - \lambda_2 + \gamma)}{s(\lambda_1 - \lambda_2)} V(\lambda_1 ) \; + \nonumber \\
&  \frac{s(\theta + \gamma -\lambda_1 + \mu_1)}{s(\theta+\gamma)} \frac{s(\lambda_1 - \mu_2 + \gamma)}{s(\mu_1 - \mu_2 + \gamma)} s(\lambda_2 - \mu_1) \frac{s(\lambda_2 - \lambda_1 + \gamma)}{s(\lambda_2 - \lambda_1)}  V(\lambda_2 ) \; , \nonumber \\
\end{align}
which can be substituted back into the original equation (\ref{L2}). This step leave us with an equation involving the functions
$V(\lambda_0 )$, $V(\lambda_1 )$, $V(\lambda_2 )$ and $V(\lambda_3 )$ for arbitrary values of those variables. 
Moreover, by setting $\lambda_3 = \mu_1$ we are left with the equation
\[
\label{L1}
K_1 V(\lambda_2) + K_2 V(\lambda_1) + L_{21} V(\lambda_0) = 0 \; ,
\] 
where the coefficients $K_1$, $K_2$ and $L_{21}$ coincide respectively with $M_1$, $M_2$ and $N_{21}$ obtained from (\ref{mm})
and (\ref{nn}) with $L=1$ ($n=2$), $\theta \rightarrow \theta + \gamma$ and $\mu_1 \rightarrow \mu_2$. So the function $V(\lambda)$ obeys the same equation as
$Z_{\theta + \gamma}(\lambda | \mu_2)$. 

We have now reached an important stage of this approach. Let us suppose that the solution of (\ref{FZ}) with polynomial structure (\ref{pol}) is unique. In fact, the uniqueness
of the polynomial solutions is demonstrated in the \appref{sec:unic}. Since $V(\lambda)$ and $Z_{\theta}(\lambda)$ are 
polynomials of the same order, that is to say $V(\lambda) = \Omega_2 Z_{\theta + \gamma}(\lambda)$ where $\Omega_2$
does not depend on $\lambda$. The solution of (\ref{FZ}) for $L=1$ can be found in the 
\appref{sec:L1} and here we shall only make use of the solution. In this way the results so far can be gathered and from (\ref{quase2}) and (\ref{sal1}) we immediately obtain an explicit solution
for $Z_{\theta}(\lambda_1 , \lambda_2)$. The solution is then given by
\[
Z_{\theta}(\lambda_1 , \lambda_2) = F_{12} + F_{21}
\]
where
\[
F_{ij} = \Omega_2 \frac{s(\theta + \gamma - \lambda_i + \mu_1)}{s(\theta + \gamma)} \frac{s(\theta + 2 \gamma - \lambda_j + \mu_2)}{s(\theta + 2 \gamma)}
\frac{s(\lambda_i - \mu_2 + \gamma)}{s(\mu_1 - \mu_2 + \gamma)} s(\lambda_j - \mu_1) \frac{s(\lambda_j - \lambda_i + \gamma)}{s(\lambda_j - \lambda_i)} \; .
\]
The constant factor $\Omega_2$ can be fixed by the asymptotic behaviour (\ref{asym}) and we find $\Omega_2 = s(\gamma)^2 s(\mu_1 - \mu_2 + \gamma)$. Thus we have completely determined
the partition function (\ref{pft}) for $L=2$ using solely the polynomial structure (\ref{pol}) and the asymptotic
behaviour (\ref{asym}), in addition to the functional equation (\ref{FZ}).
In what follows we shall consider the general $L$ case. 

\paragraph{Special Zeroes and Symmetry.} The first step in order to consider the case with arbitrary values of $L$ 
is to obtain an analogous of the relation (\ref{p2}) which can be obtained by uncovering special zeroes of our partition function.
These zeroes have been unveiled in \appref{sec:zeroes}, and in addition to that we shall also make use of the symmetry property 
discussed in \appref{sec:symmetries}. Thus taking into account (\ref{eq:zerobala}) and (\ref{eq:ssyim}) we can write
\[
\label{pL}
\left. Z_{\theta}(\lambda_1, \dots , \lambda_L ) \right|_{\lambda_1 = \mu_1} = \prod_{i=2}^{L} s(\lambda_i - \mu_1 + \gamma) V(\lambda_2 , \dots , \lambda_L ) \; .
\]
\medskip
We proceed by setting $\lambda_0 = \mu_1$ and $\lambda_{L+1} = \mu_1 - \gamma$ in the functional equation (\ref{FZ}). By doing so
we obtain the expression
\<
\label{fast1}
Z_{\theta}(\lambda_1 , \dots , \lambda_L ) =  \sum_{j=1}^{L} \prod_{\stackrel{k=1}{\neq j}}^{L} s(\lambda_k - \mu_1 + \gamma) 
\frac{\bar{m}_j}{m_L} V(\lambda_1 , \dots , \lambda_{j-1} , \lambda_{j+1}, \dots , \lambda_L ) \; , \nonumber \\
\>
where
\<
m_L &=& \frac{s(\theta + \gamma)}{s(\theta + (L+1)\gamma)} (-1)^{L+1} s(\gamma)^2 \prod_{j=2}^{L} s(\mu_1 - \mu_j + \gamma) s(\mu_j - \mu_1 + \gamma) \nonumber \\
\bar{m}_j &=&  \frac{s(\theta + \gamma - \lambda_j + \mu_1)}{s(\theta + (L+1)\gamma)} (-1)^L s(\gamma)^2 
\prod_{k=2}^{L} s(\mu_k - \mu_1 + \gamma) s(\lambda_j - \mu_k + \gamma) \nonumber \\
&& \times \; \prod_{\stackrel{k=1}{\neq j}}^{L} \frac{s(\lambda_k - \mu_1)}{s(\lambda_k - \mu_1 + \gamma)}
\frac{s(\lambda_k - \lambda_j + \gamma)}{s(\lambda_k - \lambda_j)} \; .
\>
It is important to remark here that we have also considered (\ref{pL}) and the symmetry property 
$Z_{\theta} (\dots , \lambda_i , \dots ,  \lambda_j , \dots )  = Z_{\theta} (\dots , \lambda_j , \dots ,  \lambda_i , \dots )$
discussed in the \appref{sec:symmetries} in order to obtain (\ref{fast1}).

The relation (\ref{fast1}) can now be substituted back into the Eq. (\ref{FZ}) and considering $\lambda_{L+1} = \mu_1$ we obtain
\<
\label{FZ1}
&&\sum_{i=1}^{L} K_i V(\lambda_1, \dots , \lambda_{i-1} , \lambda_{i+1} , \dots , \lambda_{L}) \nonumber \\
&& \qquad \qquad  + \sum_{j=2}^{L} \sum_{i=1}^{j-1} L_{ji} V(\lambda_0, \dots , \lambda_{i-1} , \lambda_{i+1} , \dots , \lambda_{j-1} , \lambda_{j+1} ,\dots , \lambda_{L}) = 0 \; . \nonumber \\
\>
In their turn the coefficients $K_i$ and $L_{ji}$ appearing in (\ref{FZ1}) correspond respectively to the coefficients
$M_i$ and $N_{ji}$ given in (\ref{mm}) and (\ref{nn}) with $L \rightarrow L-1$ ($n=L$), $\theta \rightarrow \theta + \gamma$
and $\mu_i \rightarrow \mu_{i+1}$. Moreover,
the compatibility between (\ref{pL}) and (\ref{pol}) tells us that the function $V$ is a multivariate polynomial of the same order as the partition function 
$Z_{\theta}$ for a lattice with dimensions $(L-1) \times (L-1)$. Thus the Eq. (\ref{FZ1}), together with the uniqueness property
discussed in the \appref{sec:unic}, implies in
\[
\label{eq:VZ}
V(\lambda_1 , \dots , \lambda_n ) = \Omega_n   Z_{\theta + \gamma}(\lambda_1 , \dots , \lambda_n ) \; .
\]
In this way the relation (\ref{fast1}) can be iterated using the results obtained in the \appref{sec:L1} as initial condition. 

By carrying on with this procedure we obtain the following expression for our partition function:
\[
\label{eq:zorro}
Z_{\theta}(\lambda_1 , \dots , \lambda_L) = \sum_{\{ i_1 , \dots , i_L \} \in \mathcal{S}_L} F_{i_1 \dots i_l}
\]
where
\begin{align}
\label{eq:forro}
& F_{i_1 \dots i_l} = \nonumber \\
& \frac{\Omega_L}{\prod_{k=2}^{L} s(\mu_1 - \mu_k + \gamma)} \prod_{n=1}^{L} \frac{s(\theta + n \gamma - \lambda_{i_n} + \mu_n)}{s(\theta + n \gamma)} \prod_{n=1}^{L} \prod_{j=n+1}^{L} s(\lambda_{i_n} - \mu_j + \gamma) \prod_{j=1}^{n-1} s(\lambda_{i_n} - \mu_j) \nonumber \\
& \times \; \prod_{n=1}^{L-1} \prod_{m>n}^{L} \frac{s(\lambda_{i_{m}} - \lambda_{i_{n}} + \gamma)}{s(\lambda_{i_{m}} - \lambda_{i_{n}})} \; .
\end{align}
Here $\mathcal{S}_L$ denotes the permutation group of order $L$ and the asymptotic behaviour (\ref{asym}) implies 
in $\Omega_L = s(\gamma)^L \prod_{k=2}^{L} s(\mu_1 - \mu_k + \gamma)$.

\section{Multiple integral representation}
\label{sec:mint}

The expression (\ref{eq:zorro}, \ref{eq:forro}) can be converted into a multiple contour integral similarly to the expression recently found in 
\cite{deGier_Galleas11} for the $U_q[\widehat{\alg{su}}(2)]$ vertex model. As a matter of fact, multiple contour integrals seems to fit naturally into
the algebraic-functional framework presented here. We start by noticing that the solution of (\ref{FZ}) for $L=1$ given in (\ref{sal1})
can be rewritten as
\[
\label{eq:intL1}
Z_{\theta} (\lambda) = \frac{s(\gamma)}{2 \pi \ii} \oint \frac{1}{s(w - \lambda)} \frac{s(\theta + \gamma - w + \mu_1)}{s(\theta + \gamma)} \dd w \; ,
\]
where the integration contour contains the pole at $w = \lambda$. Now we look to the Eq. (\ref{fast1}) considering (\ref{eq:VZ}) and keeping
in mind that for $L=1$ we have (\ref{eq:intL1}). This suggests that the iteration procedure described by (\ref{fast1}) can be mimicked by
Cauchy's residue formula. It turns out that when we look to (\ref{fast1}) searching for solutions as contour integrals, we find
a factorised formula for the integrand. In this way we end up with the following expression for our partition function,
\begin{align}
\label{eq:integral}
Z_{\theta}(\lambda_1 , \dots , \lambda_L) = & \nonumber \\
\left[ \frac{s(\gamma)}{2\pi \ii} \right]^L  \oint \dots \oint &  \frac{\prod_{i=1}^L\prod_{j=i+1}^{L} s(w_j - w_i + \gamma) s(w_j - w_i)}{\prod_{i,j=1}^L s(w_i-\lambda_j)} 
\prod_{j=1}^{L} \frac{s(\theta + j \gamma - w_{j} + \mu_j)}{s(\theta + j \gamma)} \times \nonumber \\
& \prod_{i=1}^L \prod_{j=1}^{i-1}  s(\mu_j - w_i) \prod_{j=i+1}^{L}  s(w_i - \mu_j + \gamma)\  \dd w_1 \dots \dd w_L \; ,
\end{align}
where the integration countours enclose the poles at $w_i = \lambda_j$.
As expected the expression (\ref{eq:integral}) coincides with (\ref{eq:zorro})-(\ref{eq:forro}) when evaluated using Cauchy's residue
formula. Moreover, in the limit $\theta \rightarrow \infty$ the formula (\ref{eq:integral}) reduces to the one obtained in \cite{deGier_Galleas11}
after a relabelling of the variables $\mu_j$. This relabelling does not affect the solution $Z_{\theta}(\lambda_1 , \dots , \lambda_L)$ since
this partition function is invariant under the exchange of variables $\mu_i \leftrightarrow \mu_j$ as discusssed in 
\cite{Korepin82}.

\section{Concluding remarks}
\label{sec:conclusion}

The main result of this paper is the integral representation (\ref{eq:integral}) obtained for the 
partition function of the trigonometric SOS model with domain wall boundaries. This
integral formula has been obtained by solving a functional equation derived from the dynamical 
Yang-Baxter algebra. This approach has been proposed in \cite{Galleas10,Galleas11} and here we also present
a more robust formulation of that method. 

In contrast to the case considered in \cite{Galleas10,Galleas11}, where the $\alg{su}(2)$ algebra only 
appears in the final stages of the derivation of (\ref{FZ}), here it plays an important role from the very beginning.
For instance, the derivation of (\ref{cbb}) requires the repeated use of the relations (\ref{ybsuk}) and (\ref{action}).

It is important to remark here that the elliptic version of this same partition function has been considered
previously in \cite{Pakuliak_2008, Rosengren_2008,  WengLi_2009, Tarasov_1997}. In particular, the work
\cite{Rosengren_2008} discusses the lack of a single determinant expression for this partition function 
generalising the Izergin-Korepin determinant. For the three-colouring model case, a functional equation for this partition 
function was obtained in \cite{Razumov_2009a, Razumov_2009b} though a connection with the functional equation presented here 
is not obvious at the moment. It is also worth remarking that the trigonometric SOS model with one reflecting end, and
the remaining boundaries of domain wall type has been considered in \cite{FK10, FK11}. In that case the dynamical
Yang-Baxter algebra also plays an important role, though it is only responsible for a few out of six conditions determining
uniquely the model partition function. The approach considered here makes use of only three conditions and it would
be interesting to extend it to the case considered in \cite{FK10, FK11}.

Moreover, it has been recently discussed in \cite{Colomo_2011}
the usefulness of such integral formulas for computing correlation functions for the case of domain 
wall boundaries which makes the representation (\ref{eq:integral})  more attractive. The generalisation of
our results for the elliptic case is under investigation and we hope to report on that in a future publication.

\section{Acknowledgments}
\label{sec:ack}
The author thanks J. de Gier and M. Sorrell for many useful discussions and collaboration
in \cite{deGier_Galleas11} where similar integral formulas for domain wall boundaries have appeared.
Most of the calculations presented here have been perfomed
at the Max-Planck-Institut f\"ur GravitationsPhysik (AEI) to which the author express
his sincere thanks for the excellent working conditions. Financial support from the Australian Research
Council and The Centre of Excellence for the Mathematics and Statistics of Complex Systems (\mbox{MASCOS})
is also gratefully acknowledged.

\appendix

\section{Dynamical Yang-Baxter algebra vs. $\alg{su}(2)$}
\label{sec:high}

The analysis performed here will follow the same lines as the one presented in \cite{Galleas11}. We shall
consider the $\alg{su}(2)$ generators $\gen{E}$, $\gen{F}$ and $\gen{H}$ satisfying the relations
\[
\label{su2}
[ \gen{E}, \gen{F} ] = H \qquad \qquad [ \gen{H} , \gen{E} ] = 2 \gen{E}  \qquad \qquad [ \gen{H} , \gen{F} ] = -2 \gen{F} \; ,
\]
and compute their commutation relations with the generators of the dynamical Yang-Baxter algebra
$A(\lambda,\theta)$, $B(\lambda,\theta)$, $C(\lambda,\theta)$ and $D(\lambda,\theta)$. In fact we will only need their commutation
rules with the Cartan generator $\gen{H}$ whose fundamental representation on the quantum space is given by
\[
\gen{H} = \sum_{i=1}^{L} \hat{h}_{i}  \; .
\] 
Here $\hat{h}_{i}$ consists of the Pauli matrix 
\[
\hat{h} = \left( \begin{matrix}
1 & 0 \cr
0 & -1 \end{matrix} \right)
\]
acting non-trivially on the space $\mathbb{V}_i$ of the tensor product $\mathbb{V}_1 \otimes \dots \otimes \mathbb{V}_L$.

The ice rule (\ref{ice}) can be rewritten as $[ \mathcal{R}_{aj}(\lambda,\theta_j) , \hat{h}_j ] = - [ \mathcal{R}_{aj}(\lambda,\theta_j) , \hat{h}_a ]$ 
which immediately lead us to the relation
\[
\label{tauH}
[ \mathcal{T}_a (\lambda, \theta) , \gen{H} ] = - [ \mathcal{T}_a (\lambda, \theta) , \hat{h}_a ]
\]
due to the definition (\ref{mono}). In terms of the monodromy matrix entries (\ref{abcd}), the relation (\ref{tauH}) explicitly reads
\begin{align}
\label{ybsuapp}
\left[ A(\lambda, \theta) , \gen{H} \right] &= 0 &  \left[ B(\lambda, \theta) , \gen{H} \right] &= 2  B(\lambda, \theta) \nonumber \\
\left[ C(\lambda, \theta) , \gen{H} \right] &= -2 C(\lambda, \theta)&  \left[ D(\lambda, \theta) , \gen{H} \right] &= 0 \; ,
\end{align}
which allows us to exploit the representation theory of the $\alg{su}(2)$ algebra in order to gain insight into the dynamical Yang-Baxter algebra generators.

For instance, the $\alg{su}(2)$ highest and lowest weight states $\ket{0}$ and $\ket{\bar{0}}$ defined in (\ref{states1}) obey the
relations $\gen{H} \ket{0} = L \ket{0}$ and $\gen{H} \ket{\bar{0}} = -L \ket{\bar{0}}$. These properties
together with (\ref{ybsuapp}) allow us to obtain the relation
\[
\label{h0}
\gen{H} \prod_{i=1}^{n} B(\lambda_i, \theta + (i-1)\gamma) \ket{0} = (L - 2n) \prod_{i=1}^{n} B(\lambda_i, \theta + (i-1)\gamma) \ket{0}
\]
which is valid for any number $n$ of operators $B(\lambda, \theta)$. Now the relation (\ref{h0}) put us in position to use the $\alg{su}(2)$ representation
theory to draw conclusions about the generator $B(\lambda, \theta)$. For the case $n=L$ the expression (\ref{h0}) tells us that 
$\prod_{i=1}^{L} B(\lambda_i, \theta + (i-1)\gamma) \ket{0}$ is an eigenvector of $\gen{H}$ with eigenvalue $-L$. On the other hand this is the same
eigenvalue associated with the state $\ket{\bar{0}}$. Since this eigenvalue is not degenerated we can conclude that
\[
\prod_{i=1}^{L} B(\lambda_i, \theta + (i-1)\gamma) \ket{0} \sim \ket{\bar{0}} \; ,
\] 
and from (\ref{destr}) we immediately have that
\[
\label{vas}
\prod_{i=1}^{L+1} B(\lambda_i, \theta + (i-1)\gamma) \ket{0} = 0 \; .
\]
The property (\ref{vas}) is an important ingredient for the derivation of the functional equation (\ref{FZ}).

\section{Polynomial structure and asymptotic behaviour}
\label{sec:polstruc}

In order to analyse the dependence of $Z_{\theta}$ with the set of variables $\{  \lambda_i  \}$
we first consider the following change of variables:
\begin{align}
x_i &= e^{2 \lambda_i} & \quad u_i &= e^{2 \mu_i} & \quad q &= e^{\gamma} \nonumber \\
\bar{x}_i &= e^{\lambda_i}&  \quad \bar{u}_i &= e^{\mu_i} & \quad t &= e^{\theta} \; . 
\end{align}
In terms of the above defined variables,  the $\mathcal{R}$-matrix given by (\ref{rmat}) and (\ref{bw}) can be written as
\[
\label{bax}
\mathcal{R} = \frac{1}{8 q \bar{x}} \left( x \gen{U} + \gen{V} \right) \; ,
\]
where
\<
\label{UV}
\gen{U} &=& (3 q^2 +1) \gen{1} \otimes \gen{1} + (q^2 -1) \gen{H} \otimes \gen{H} + \frac{(q^2 -1)(t^2 +1)}{(t^2-1)} \gen{1} \otimes \gen{H}
-  \frac{(q^2 -1)(t^2 +1)}{(t^2-1)} \gen{H} \otimes \gen{1} \nonumber \\ 
&+& \frac{4(1-q^2)}{(t^2 -1)} \gen{E} \otimes \gen{F} + \frac{4 t^2 (q^2 -1)}{(t^2 -1)} \gen{F} \otimes \gen{E} \nonumber \\
\gen{V} &=& -(3 + q^2) \gen{1} \otimes \gen{1} + (q^2 -1) \gen{H} \otimes \gen{H} - \frac{(q^2 -1)(t^2 +1)}{(t^2-1)} \gen{1} \otimes \gen{H}
+  \frac{(q^2 -1)(t^2 +1)}{(t^2-1)} \gen{H} \otimes \gen{1} \nonumber \\ 
&+& \frac{4 t^2 (q^2 -1)}{(t^2 -1)} \gen{E} \otimes \gen{F} + \frac{4 (1 - q^2)}{(t^2 -1)} \gen{F} \otimes \gen{E} \; .
\>
In (\ref{UV}) the generators $\gen{E}$, $\gen{F}$ and $\gen{H}$ are the $\alg{su}(2)$ generators satisfying (\ref{su2}),
and considering (\ref{bax}), (\ref{dmono}) and (\ref{abcd}) we readly obtain the expansion
\[
B(\lambda_i , \theta) = \frac{1}{\bar{x}^{L}_i} \left[ f^{(i)}_{L} x^{L}_i + f^{(i)}_{L-1} x^{L-1}_i + \dots +f^{(i)}_{0}  \right] \; .
\]

Now looking to the product  $\prod_{j=1}^{L} B(\lambda_j, \theta + j \gamma)$ appearing in the definition (\ref{pft}), we can conclude that
\[
Z_{\theta}(\lambda_1 , \dots , \lambda_L) = \frac{\bar{Z}_{\theta} (x_1 , \dots , x_L)}{\displaystyle \prod_{i=1}^{L} \bar{x}^{L}_i} \; ,
\]
where $\bar{Z}_{\theta} (x_1 , \dots , x_L)$ is a polynomial of order $L$ in each variable $x_i$. 

Also from (\ref{bax})  we can see that in the limit $x \rightarrow \infty$ only the operator $\gen{U}$ contributes for the partition
function $Z_{\theta}$. In (\ref{UV}) the operator $\gen{U}$ is written in terms of $\gen{su}(2)$ generators which allows us
to follow the same analysis of \cite{Galleas10}. Without significant modifications we find that
\[
\bar{Z} \sim \frac{(q-q^{-1})^L}{2^{L^2}} \frac{[ L ]_{q^2} ! }{\displaystyle \prod_{n=1}^{L} (1 - q^{2n} t^2 ) \bar{u}_{n}^{L}}  (x_1 \dots x_L)^L  
\quad \mbox{as} \quad \; x_i \rightarrow \infty \; .
\]
Here the $q$-factorial function is defined as
\[ 
[ n ]_q ! = 1 (1+q)(1+q+q^2) \dots (1+q+ \dots +q^n) \; .
\]

\section{Special Zeroes}
\label{sec:zeroes}

One important ingredient for solving the Eq. (\ref{FZ}) under the conditions 
(\ref{pol}) and (\ref{asym}) is the localisation of some special zeroes of our
partition function. Since we are interested in the polynomial solution of (\ref{FZ}),
those zeroes will play an important role in the characterisation of our solution. We shall start
by looking to particular values of $L$ for illustrative purposes and then we proceed to the general 
case. 

\begin{itemize}
\item {\bf $L=2$:}
\end{itemize}
We set $\lambda_3 = \mu_1$ and $\lambda_2 = \mu_1-\gamma$ in such a way that the functions
$M_2$, $M_3$, $N_{21}$, $N_{31}$ and $N_{32}$ vanish. For these particular values of $\lambda_3$
and $\lambda_2$ we are thus left with
\[
\left. M_1 \right|_{\lambda_2 , \lambda_3} Z_{\theta} (\mu_1-\gamma , \mu_1) = 0 \; .
\]
Since $\left. M_1 \right|_{\lambda_2, \lambda_3 }$ is finite we can conclude that
$Z_{\theta} (\mu_1-\gamma , \mu_1) = 0$ .

\begin{itemize}
\item {\bf $L=3$:}
\end{itemize}
By setting $\lambda_4 = \mu_1$ and $\lambda_3 = \mu_1-\gamma$ we obtain
\[
\sum_{i=0}^{2} P_i \; Z_{\theta} (\mu_1 , \mu_1 - \gamma , \lambda_i) = 0
\]
where
\<
P_0 = \left. N_{21} \right|_{\lambda_3 , \lambda_4} \quad , \quad P_1 = \left. M_{2} \right|_{\lambda_3 , \lambda_4} 
\quad \mbox{and} \quad P_2 = \left. M_{1} \right|_{\lambda_3 , \lambda_4} \; .
\>
In terms of the variables $x_i$, the functions $P_i$ are rational functions and thus
$\exists \; \lambda_i : P_i = 0$. Besides the above specialisation of the variables
$\lambda_4$ and $\lambda_3$, we also choose $\lambda_i \; | \; P_i = 0$ for $i=1,2$. Thus we are left with
\[
\left. P_{0} \right|_{\lambda_1 , \lambda_2}  Z_{\theta} (\mu_1 , \mu_1 - \gamma , \lambda_0) = 0 \; ,
\]
and since $\left. P_{0} \right|_{\lambda_1 , \lambda_2}$ is finite we can conclude that 
$Z_{\theta} (\mu_1 , \mu_1 - \gamma , \lambda_0) = 0$.

\begin{itemize}
\item {\bf $L=4$:}
\end{itemize}
For the case $L \geq 4$ this analysis becomes a bit more involved. We start by setting $\lambda_5 = \mu_1$ and $\lambda_4 = \mu_1-\gamma$
similarly to the previous cases. Under this specialisation the Eq. (\ref{FZ}) reduces to
\begin{align}
\label{lala}
& \left. M_1 \right|_{\lambda_4 , \lambda_5} Z_{\theta}(\lambda_2 , \lambda_3 , \mu_1 - \gamma , \mu_1) + \left. M_2 \right|_{\lambda_4 , \lambda_5} Z_{\theta}(\lambda_1 , \lambda_3 , \mu_1 - \gamma , \mu_1) \; +  \nonumber \\             
& \left. M_3 \right|_{\lambda_4 , \lambda_5} Z_{\theta}(\lambda_1 , \lambda_2 , \mu_1 - \gamma , \mu_1) + \left. N_{21} \right|_{\lambda_4 , \lambda_5} Z_{\theta}(\lambda_0 , \lambda_3 , \mu_1 - \gamma , \mu_1) \; +  \nonumber \\ 
& \left. N_{31} \right|_{\lambda_4 , \lambda_5} Z_{\theta}(\lambda_0 , \lambda_2 , \mu_1 - \gamma , \mu_1) + \left. N_{32} \right|_{\lambda_4 , \lambda_5} Z_{\theta}(\lambda_0 , \lambda_1 , \mu_1 - \gamma , \mu_1) = 0 \; .
\end{align}
Next we set $\lambda_0 = \mu_1$ and $\lambda_1 = \mu_1-\gamma$. The Eq. (\ref{lala}) does not suffer
significant simplifications and we then proceed by setting $\lambda_3 = \mu_1$ using the following properties:
\<
\lim_{\lambda_3 \rightarrow \mu_1} \left. \frac{M_1}{N_{31}} \right|_{\lambda_0 , \lambda_1 , \lambda_4 , \lambda_5} &=& - 1 \nonumber \\
\lim_{\lambda_3 \rightarrow \mu_1} \left. \frac{M_2}{N_{32}} \right|_{\lambda_0 , \lambda_1 , \lambda_4 , \lambda_5} &=& - 1 \; .
\>
By doing so we end up with the relation
\[
\label{lele}
\left. M_3 \right|_{\lambda_0, \lambda_1, \lambda_3, \lambda_4, \lambda_5 } Z_{\theta} (\mu_1 - \gamma, \lambda_2 , \mu_1 - \gamma, \mu_1) =
\left. N_{21} \right|_{\lambda_0, \lambda_1, \lambda_3, \lambda_4, \lambda_5 } Z_{\theta} (\mu_1, \mu_1 , \mu_1 - \gamma, \mu_1) \; ,
\] 
which is further simplified to
\[
\label{lili}
\left. N_{21} \right|_{\lambda_0, \lambda_1, \lambda_2 ,\lambda_3, \lambda_4, \lambda_5 } Z_{\theta} (\mu_1, \mu_1 , \mu_1 - \gamma, \mu_1) = 0
\]
with $\lambda_2 = \mu_1 - \gamma$. The quantity $\left. N_{21} \right|_{\lambda_0, \lambda_1, \lambda_2 ,\lambda_3, \lambda_4, \lambda_5 }$
is finite which allow us to conclude that
\[
\label{lolo}
Z_{\theta} (\mu_1, \mu_1 , \mu_1 - \gamma, \mu_1) = 0 \; .
\]

Now we move backwards considering the consequences of (\ref{lolo}) to the previous equations. From (\ref{lili}) and (\ref{lolo})
we have that
\[
\label{lulu}
Z_{\theta} (\mu_1 - \gamma, \lambda_2 , \mu_1 - \gamma, \mu_1) = 0 \; ,
\]
which can be reintroduced in (\ref{lala}) with the above mentioned specialisation of $\lambda_0$ and $\lambda_1$. This yields
the following expression
\<
\label{mama}
Z_{\theta} (\lambda_2, \lambda_3 , \mu_1 - \gamma, \mu_1) = &-& \left. \frac{N_{21}}{M_1} \right|_{\lambda_0 , \lambda_1 , \lambda_4 , \lambda_5} Z_{\theta} (\mu_1, \lambda_3 , \mu_1 - \gamma, \mu_1) \nonumber \\
&-& \left. \frac{N_{31}}{M_1} \right|_{\lambda_0 , \lambda_1 , \lambda_4 , \lambda_5} Z_{\theta} (\mu_1, \lambda_2 , \mu_1 - \gamma, \mu_1) \; .
\>
Now we replace (\ref{mama}) back into (\ref{lala}) to obtain the expression
\[
\sum_{i=0}^{3} Q_i \; Z_{\theta} (\mu_1 , \lambda_i , \mu_1 - \gamma , \mu_1) = 0 \; . 
\]
The explicit form of $Q_i$ is not enlightening and  shall not be presented here. Nevertheless, using
similar arguements as for the cases $L=2,3$ we can conclude that $Z_{\theta} (\mu_1 , \lambda , \mu_1 - \gamma , \mu_1) = 0$.
Thus from (\ref{mama}) we obtain the vanishing condition 
\[
\label{eq:zerobala}
Z_{\theta} (\lambda_2, \lambda_3 , \mu_1 - \gamma, \mu_1) = 0 \; .
\]

\begin{itemize}
\item {\bf General $L$:}
\end{itemize}
For arbitrary values of $L$ we initially set $\lambda_{L+1} = \mu_1$ and $\lambda_{L} = \mu_1 - \gamma$ in the functional
equation (\ref{FZ}), followed by the specialisation $\lambda_{0} = \mu_1$ and $\lambda_{1} = \mu_1 - \gamma$. We collect the results at  
each one of the steps and then start fixing the variables $\lambda_{L-1} = \mu_1$, $\lambda_{L-2} = \mu_1-\gamma$, $\lambda_{L-3} = \mu_1$
and so on until we exhaust all the variables. Then the consistency condition of each step with the previous ones allow us to conclude that
\[
\label{meme}
Z_{\theta} (\mu_1 , \mu_1 - \gamma , \lambda_3 , \dots , \lambda_L   ) = 0 \; .
\]
Together with the symmetry property discussed in \appref{sec:symmetries}, the relation (\ref{meme}) plays an important
role for solving (\ref{FZ}).

\section{$Z_{\theta}$ as a symmetric function}
\label{sec:symmetries}

In this appendix we intend to show that Eq. (\ref{FZ}) admits only analytic solutions
which are symmetric under the exchange of variables $\lambda_i \leftrightarrow \lambda_j$. This is an
expected property of our partition function (\ref{pft}) due to the commutation relations (\ref{commut}).
Nevertheless, we shall demonstrate that this property is not an extra input required to solve Eq. (\ref{FZ}). 

We start by integrating the Eq. (\ref{FZ}) over the contour $\mathcal{C}_j$ containing only the variable $\lambda_j$.
For a given $j$, the coefficients $M_j$ and $N_{kl}$ $(k,l \neq j)$ do not contain poles when $\lambda_0 \rightarrow \lambda_j$.
Moreover, we also have the following identities between the coefficients
\<
\label{id}
\lim_{\lambda_0 \to \lambda_j} s(\lambda_0 - \lambda_j) \; M_k &=& - \lim_{\lambda_0 \to \lambda_j} s(\lambda_0 - \lambda_j) \;  N_{jk} \qquad \qquad k<j \nonumber \\
\lim_{\lambda_0 \to \lambda_j} s(\lambda_0 - \lambda_j) \; M_k &=& - \lim_{\lambda_0 \to \lambda_j} s(\lambda_0 - \lambda_j) \;  N_{kj} \qquad \qquad k>j 
\> 
for $j=1 , \dots , L$. Thus after the integration of (\ref{FZ}) over the contour $\mathcal{C}_j$, we are left with the relation
\<
\label{id1}
\sum_{\stackrel{i=1}{\neq j}}^{L} \frac{\check{M}_i}{\check{M}_{L+1}} && \left[ Z_{\theta}(\lambda_1 , \dots , \lambda_{i-1}, \lambda_{i+1}, \dots , \lambda_{L+1} ) 
- Z_{\theta}(\lambda_j , \lambda_1 , \dots , \lambda_{i-1}, \lambda_{i+1}, \dots , \lambda_{L+1} ) \right] = \nonumber \\ 
&& Z_{\theta}(\lambda_j , \lambda_1 , \dots , \lambda_{L} ) - Z_{\theta}(\lambda_1 , \dots , \lambda_{L})
\>
where
\[
\check{M}_k =  \lim_{\lambda_0 \to \lambda_j} s(\lambda_0 - \lambda_j) \;  M_k \; .
\]

Two observations are important at this stage. Firstly, the relation (\ref{id1}) is valid for
$j=1 , \dots , L$ and thus it provides us with a total of $L$ equations. Secondly, 
we notice that the RHS of (\ref{id1}) does not depend on the variable $\lambda_{L+1}$. In fact
this variable can be adjusted, together with the results obtained for the $(j-1)$-th equation, in 
order to show that the RHS of (\ref{id1}) vanishes. Thus the relation (\ref{id1}) implies in
\[
Z_{\theta}(\lambda_1 , \dots , \lambda_L) = Z_{\theta}(\lambda_j , \lambda_1 ,  \dots , \lambda_{j-1} , \lambda_{j+1} , \dots,  \lambda_L)
\]
for $j=1 , \dots , L$ and consequently we have the desired symmetry relation
\[
\label{eq:ssyim}
Z_{\theta}(\lambda_1 , \dots , \lambda_i , \dots , \lambda_j , \dots , \lambda_L) = 
Z_{\theta}(\lambda_1 , \dots , \lambda_j , \dots , \lambda_i , \dots , \lambda_L) \; .
\]

\section{Uniqueness}
\label{sec:unic}

In this appendix we  prove the uniqueness of the multivariate polynomial solution 
of the Eq. (\ref{FZ}). In order to start we first need to introduce the modified coefficients 
\[
\bar{M}_i = \prod_{\stackrel{j=1}{\neq i}}^{L+1} x_{j}^{-\frac{L}{2}} M_i  \qquad \mbox{and} \qquad \bar{N}_{ji} = \prod_{\stackrel{k=0}{\neq i,j}}^{L+1} x_{j}^{-\frac{L}{2}} N_{ji} \; , 
\]
with $x_i = e^{2 \lambda_i}$. In this way the Eq. (\ref{FZ}) is given by
\<
\label{mFZ}
&&\sum_{i=1}^{L+1} \bar{M}_i \bar{Z}_{\theta}(\lambda_1, \dots , \lambda_{i-1} , \lambda_{i+1} , \dots , \lambda_{L+1}) \nonumber \\
&& \qquad \qquad  + \sum_{j=2}^{L+1} \sum_{i=1}^{j-1} \bar{N}_{ji} \bar{Z}_{\theta}(\lambda_0, \dots , \lambda_{i-1} , \lambda_{i+1} , \dots , \lambda_{j-1} , \lambda_{j+1} ,\dots , \lambda_{L+1}) = 0 \; , \nonumber \\
\>
in terms of $\bar{Z}_{\theta}$ which is a polynomial of order $L$ in each variable $x_i$ according to (\ref{pol}).

We shall now explore the linearity of the Eq. (\ref{mFZ}). More precisely that means the following: if $\bar{Z}_1$ and $\bar{Z}_2$ are two multivariate polynomials
of type (\ref{pol}) satisfying (\ref{mFZ}), then 
\[
\label{linear}
\bar{Z} = \alpha \bar{Z}_1 - \beta \bar{Z}_2
\]
is also a solution for any constants $\alpha$ and $\beta$. Polynomials are characterised by the location of their zeroes and we can express $\bar{Z}$, $\bar{Z}_1$ and $\bar{Z}_2$ as
\[
\label{lpol}
\bar{Z} \sim \prod_{i=1}^{L} (x - r_i)  \qquad \bar{Z}_1 \sim \prod_{i=1}^{L} (x - s_i) \qquad \bar{Z}_2 \sim \prod_{i=1}^{L} (x - t_i)  
\]
where $x$ can represent any of the variables $x_i$. Since the constants $\alpha$ and $\beta$ in (\ref{linear}) are arbitrary they can always be fine tunned
in order to ensure that $\bar{Z}$ is also of order $L$. Next we set $x = r_j$ in (\ref{linear}) and from (\ref{lpol}) we obtain
\[
\label{cond}
\alpha \prod_{i=1}^{L} (r_j - s_i) \sim \beta \prod_{i=1}^{L} (r_j - t_i) \; .
\]
The relation (\ref{cond}) allows us to make important conclusions. For instance, if we assume that $\{ r_i \} \neq \{ s_i \}$ then (\ref{cond})
implies that $\{ s_i \} = \{ t_i \}$ since $\alpha$ and $\beta$ can always be adjusted to compensate an overall factor. This implies that
$\bar{Z}_1$ and $\bar{Z}_2$ are proportional to each other and so is $\bar{Z}$ due to (\ref{linear}). This consequence clearly contradicts the initial
assumption $\{ r_i \} \neq \{ s_i \}$. The remaining option is allowing $\{ r_i \} = \{ s_i \}$ and thus $\bar{Z}$ and $\bar{Z}_1$ only differ by a constant. 
This fact together with (\ref{linear}) tell us that $\bar{Z}_2$ is also proportional to $\bar{Z}_1$. In summary this analysis shows that if we have two polynomials
of the same order solving (\ref{mFZ}), they are essentially the same polynomial. This proves the uniqueness of the polynomial solution 
of (\ref{FZ}).

\section{Solution for $L=1$}
\label{sec:L1}

Here we shall present the solution of the Eq. (\ref{FZ}) for the case $L=1$ which is of fundamental importance in order to derive
the solution for general $L$. For the case $L=1$ the Eq. (\ref{FZ}) reads
\[
\label{LL1}
M_1 Z_{\theta} (\lambda_2) + M_2 Z_{\theta} (\lambda_1) + N_{21} Z_{\theta}(\lambda_0) = 0 \; .
\]
At first look the condition $\lambda_i = \lambda$ does not seem helpful in finding the solution
of (\ref{LL1}). However, a closer look reveals that the coefficients $M_1$, $M_2$ and $N_{21}$ contain poles when
the variables $\lambda_i$ coincide. As we shall see this fact will be of fundamental importance. 

We can compute the limit $\lambda_i = \lambda$ of the Eq. (\ref{LL1}) using L'Hopital's rule
and we are left with the following second order differential equation
\[
\label{DZ}
P_0 \bar{Z}_{\theta} + P_1 \frac{d \bar{Z}_{\theta}}{dx} + P_2 \frac{d^2 \bar{Z}_{\theta}}{dx^2} = 0 \; ,
\]
given in terms of variables $x=e^{2 \lambda}$ and $u=e^{2 \mu_1}$. The coefficients in (\ref{DZ}) are given by
\<
\label{p0p1p2}
P_0 &=& (-4 q^2+2 q^4 t^2+2 q^6 t^2 )u + (2 q^2+2 q^4-4 q^6 t^2) x \nonumber \\
P_1 &=& (-4 q^4 t^2+2 q^6 t^4+2 q^8 t^4 ) u^2 + (4 q^2 -4 q^8 t^4 ) x u + (-2 q^2-2 q^4+4 q^6 t^2) x^2 \nonumber \\
P_2 &=& (1+q^2 -4 q^4 t^2 +q^6 t^4 +q^8 t^4  )x u^2 + (-4 q^2 -q^2 t^2 +5 q^4 t^2 +5 q^6 t^2 -q^8 t^2 -4 q^8 t^4 )u x^2 \nonumber \\
&+& (q^2+q^4-4 q^6 t^2+q^8 t^4+q^{10} t^4) x^3 \; ,
\>
and by standard methods we find the general solution
\[
\label{sal}
\bar{Z}_{\theta} (x) = C_1  (x - q^2 t^2 u) + C_2 (x - q^2 t^2 u) \int \frac{e^{-\int^{x} \frac{P_1 (x')}{P_2 (x')} dx'}}{(x - q^2 t^2 u)} dx \; ,
\]
where $C_1$ and $C_2$ are two arbitrary integration constants. Now the polynomial structure (\ref{pol}) asks for $C_2 = 0$, while
the asymptotic behaviour (\ref{asym}) implies in $C_1 =  \frac{(q-q^{-1})}{2} (1 - t^2 q^2)^{-1} u_{1}^{-\frac{1}{2}}$.
Thus our partition function for $L=1$ is given by
\[
\label{sal1}
Z_{\theta}(\lambda) = s(\gamma) \frac{s(\theta + \gamma -\lambda + \mu_1)}{s(\theta + \gamma)} \; .
\]

\bibliographystyle{hunsrt}
\bibliography{references}

\end{document}